\documentclass[11pt,onecolumn,floatfix,altaffilletter,superscriptaddress,tightenlines,showpacs,showkeys,preprintnumbers,nofootinbib]{revtex4-2}
\pdfoutput=1
\usepackage[colorlinks=true,citecolor=blue,linkcolor=blue,breaklinks=true]{hyperref}
\usepackage{amsmath,amssymb}
\usepackage{epsfig} 
\usepackage{graphicx,wasysym}        
\usepackage{url}
\usepackage{color}
\usepackage{multirow}
\usepackage{placeins}
\usepackage[dvipsnames]{xcolor}
\usepackage{braket}
\usepackage{float}
\usepackage{slashed,comment}
\usepackage{booktabs}
\hypersetup{
  colorlinks=true,
  linkcolor=red,
  filecolor=magenta,   
  urlcolor=violet,
  citecolor=blue,
}

\renewcommand{\arraystretch}{1.5}
\hypersetup{colorlinks,linkcolor={blue},citecolor={teal},urlcolor={violet}}

\allowdisplaybreaks

\bibpunct{[}{]}{,}{n}{}{,}

\def\beq{\begin{equation}}
\def\eeq{\end{equation}}
\def\bea{\begin{eqnarray}}
\def\eea{\end{eqnarray}}
\makeatletter
\@addtoreset{equation}{section}

\newcommand{\NNU}{Department of Physics and Institute of Theoretical Physics,
Nanjing Normal University, Nanjing, 210023, China}
\newcommand{\keyL}{Nanjing Key Laboratory of Particle Physics and Astrophysics, Nanjing 210023, China}
\begin{document}
\begin{flushleft}
CPTNP-2025-041
\end{flushleft}
\title{Probing thermal leptogenesis and dark matter through primordial gravitational waves from a supercooled universe}
   \author{Peter Athron}
   \email{peter.athron@njnu.edu.cn}
   \affiliation{\NNU}
   \affiliation{\keyL}
   \author{Satyabrata Datta}
   \email{amisatyabrata703@gmail.com}
   \affiliation{\NNU}
   \affiliation{\keyL}
   \author{Zhao-Yang Zhang}
   \email{zhaoyangzhang008@outlook.com}
   \affiliation{\NNU}
  
\begin{abstract}
We explore the cosmological dynamics of a supercooled first-order phase transition in the classically conformal $U(1)_{B-L}$ extension of the Standard Model, where radiative symmetry breaking simultaneously generates the right-handed neutrino (RHN) masses, and a strong stochastic gravitational-wave (GW) background. The slow decay of the scalar field into RHNs can induce an early matter-dominated (EMD) era whose duration is sensitive to the RHN mass and gauge coupling $g^\prime$. This non-standard cosmological phase reshapes the GW spectrum and leaves a distinctive RHN-mass-dependent spectral distortion that correlates with the flavour regime of thermal leptogenesis. Within this framework, one RHN can serve as a dark matter candidate produced nonthermally from scalar decays, while the remaining states generate the baryon asymmetry via thermal leptogenesis. For $g^\prime=0.5$, we identify such a parameter region, and show that with singlet extensions, even with a smaller gauge coupling, one can realise this mechanism for the three-flavour regime. The resulting GW signals, amplified by supercooling and modified by EMD, provide a unique window to probe the scale and flavour structure of leptogenesis in future high frequency GW observations.

\end{abstract}
\maketitle
\newpage
\section{Introduction}

The origin of the baryon asymmetry of the Universe (BAU), the nature of dark matter (DM), and the generation of neutrino masses are among the most profound questions in modern cosmology and particle physics \cite{ParticleDataGroup:2020ssz,Planck:2018vyg,Weinberg:1979bt,Kolb:1979qa,Fukugita:1986hr}. Despite the success of the Standard Model (SM), it fails to account for these phenomena, suggesting the presence of new physics beyond the Standard Model (BSM). Among the many possible extensions, the conformal scale-invariant models have emerged as a compelling candidate \cite{Iso:2009ss,Okada:2012sg,Hambye:2013dgv,khoze:2013uia,Khoze:2013oga,Lindner:2014oea,Humbert:2015epa,Oda:2015gna,Das:2016zue,Das:2015nwk,Mohapatra:2023aei}. Conformal models avoid the introduction of explicit masses to the Lagrangian, and instead dynamically generate all mass scales from quantum effects, elegantly avoiding the need to introduce ad-hoc mass scales by hand. In doing so, they also offer a possible route to address the gauge hierarchy problem in particle physics, while providing a unified framework for addressing fundamental puzzles in cosmology.

An intriguing phenomenological consequence of such frameworks is the occurrence of strong first-order phase transitions (FOPTs) \cite{Guth:1980zk,Witten:1980ez,Iso:2017uuu,Bian:2019szo, Marzo:2018nov} induced by radiative symmetry breaking \cite{Coleman:1973jx}. In these scenarios, the conformal symmetry of a BSM scalar is spontaneously broken at high temperatures, giving rise to a large potential barrier between metastable and true vacua. This results in a brief vacuum domination and makes the transition delayed or supercooled, releasing a significant amount of latent heat as strong stochastic gravitational-wave background (SGWB) \cite{Konstandin:2017sat,Jaeckel:2016jlh,vonHarling:2017yew,Hambye:2018qjv,Baldes:2018emh, Prokopec:2018tnq,Kierkla:2023von,Jinno:2016knw,Marzola:2017jzl,Marzo:2018nov,Mohamadnejad:2019vzg,Lewicki:2021xku,Lewicki:2024sfw,Goncalves:2025uwh,Schmitt:2024pby,Li:2025nja,Costa:2025csj}. These gravitational waves (GWs) carry direct information about the microphysics of the transition and can serve as cosmological relics of the conformal dynamics. 

The conformal $U(1)_{B-L}$ extension of the SM provides a particularly simple and predictive realisation of this idea \cite{Iso:2009ss,Khoze:2013oga,Bian:2019szo}. Once the $B-L$ symmetry is broken, right-handed neutrinos (RHNs) acquire their Majorana masses through the vacuum expectation value (VEV) of the $B-L$ scalar, naturally implementing the type-I seesaw mechanism. The out-of-equilibrium decays of these heavy RHNs can generate a lepton asymmetry that is partially converted into the observed BAU by electroweak sphalerons \cite{Fukugita:1986hr, Pilaftsis:2003gt,Buchmuller:2004nz,Davidson:2008bu,Riotto:1999yt}. The same particle content can also accommodate a viable DM candidate if one of the RHNs is cosmologically stable, leading to an economical and unified picture of neutrino masses, baryogenesis, and DM within a minimal extension of the SM \cite{uni1,uni2,uni3,uni4,uni5,uni6,uni7}.

A remarkable feature of the conformal $B-L$ phase transition is not only the enhanced GW signature, but also a peculiar thermal history after the completion of the phase transition \cite{Ellis:2019oqb,Ellis:2020nnr,Gouttenoire:2023pxh,Gonstal:2025qky}. The reheating after the transition can proceed inefficiently through the decay of the scalar field into RHNs, leading to an early matter-dominated (EMD) epoch before the Universe becomes radiation dominated again \cite{Blasi:2020wpy,Datta:2022tab,Datta:2023vbs}. This non-standard cosmological phase has two crucial consequences. First, it dilutes the relic abundances of pre-existing species, including DM and any prior asymmetry. Second, it imprints distinctive features on the SGWB spectrum, potentially shifting the peak frequency and modifying the spectral slopes across different frequency bands \cite{Ellis:2020nnr,Gouttenoire:2023pxh}. The interplay between leptogenesis, DM production, and GW signatures in such non-standard cosmologies provides a unique probe of high-scale conformal dynamics that would otherwise remain experimentally inaccessible. 

In this work, we investigate the cosmological implications of a supercooled FOPT in the classically conformal $B-L$ model.\footnote{While we were preparing this work for publication, we became aware of related work presented in Ref.\ \cite{2974136}.} We focus on the post-transition evolution where the scalar field decays into RHNs, inducing an EMD phase whose duration depends on the RHN mass and decay width. We analyze how this phase affects the relic abundance of a RHN DM candidate and delineate the parameter space consistent with successful thermal leptogenesis and the observed DM density. Furthermore, we compute the resulting GW spectra, emphasising the modifications induced by the EMD epoch and their potential detectability across multiple frequency bands. Also, an indirect connection between the flavour regime of leptogenesis and the GW signal can offer an exciting opportunity to indirectly probe the scale and flavour structure of leptogenesis through future GW observations.

The rest of the paper is organised as follows. In Sec.\ \ref{s2}, we briefly discuss the basic framework of supercooling in the conformal $B-L$ model. In Sec.\ \ref{s3}, we discuss how the RHN mass impacts the scalar-induced EMD after the phase transition. Sec.\ \ref{s4} contains a discussion of thermal leptogenesis and dark matter from scalar decay. In Sec.\ \ref{s5} we discuss the modification of GW spectral shapes due to an EMD. In Sec.\ \ref{s6}, we present our numerical results and delineate the allowed parameter space compatible with thermal leptogenesis and RHN DM. Finally, we conclude in Sec.\ \ref{s7}.
 
\section{The Scale-invariant Conformal $(B-L)$ Model \& supercooled FOPT}\label{s2}
We consider the conformal $(B-L)$ extension of the Standard Model augmented by a scalar singlet $\Phi=(\phi+i\eta)/\sqrt{2}$ with a $B-L$ charge of $+2$, which spontaneously breaks $U(1)_{B-L}$ and is responsible for generating the mass of RHNs with $B-L$ charge $-1$ \cite{Iso:2009nw,Iso:2009ss,Das:2015nwk}(see Table \ref{table:BLcharges} for details). It is necessary to introduce three generations of RHNs for gauge anomaly cancellation. Among them, two of the quasi-degenerate RHNs ($N_i$) are responsible for neutrino mass generation and leptogenesis. While the lightest RHN ($N_{\rm DM}$) may serve as a candidate for DM due to its feeble coupling with the Standard Model. 

The key part of the Lagrangian relevant to PT dynamics and leptogenesis is,
\begin{equation}\label{Leff}
    \mathcal{L}_{\rm int}\supset -\frac{1}{2}\sum_{i=1,2}\lambda_{Ri}\Phi \bar{N_i}N_i -\frac{1}{2}\lambda_{\rm DM}\Phi \bar{N}_{\rm DM}N_{\rm DM} -\sum_{i=1,2} Y_{ij}^\nu \bar{\ell_i} \tilde{H} N_j -V(\Phi) + {\rm h.c. },
\end{equation}
where we assume a common Yukawa coupling $\lambda_R \equiv \lambda_{R_{1,2}}$ for $N_{1,2}$, which then implies equal masses $M_{N} \equiv M_{N_{1,2}}  =\frac{\lambda_{R}}{\sqrt{2}}v_\Phi$.  Similarly the DM candidate gets a mass $M_{\rm DM}=\frac{\lambda_{\rm DM}}{\sqrt{2}}v_\Phi$ with vacuum expectation value (VEV) $v_\Phi$. Moreover, we will be interested in scenarios where the dark matter is much lighter than the other RHNs, so that the scalar field dynamics after the FOPT depend only on the mass scale of leptogenesis ($M_N$), and thus we assume $\lambda_{\rm DM}\ll\lambda_R$.   The conformal scale invariant potential at tree level can be written as,
\begin{equation}
    V_{\rm tree}(\Phi)=\lambda_\phi |\Phi|^4, 
\end{equation}
where we have assumed that the $B-L$ scale significantly exceeds the EW scale, rendering the Higgs-portal coupling negligible for radiatively generating the SM Higgs mass and effectively making the PT dynamics nearly decoupled. The 1-loop contributions from relevant bosons ($Z^\prime$) and fermions ($N_i, N_{\rm DM}$) induce a Coleman-Weinberg effective potential for $\Phi$. We compute this in the unitary gauge and following Ref.\ \cite{Coleman:1973jx} we resum leading logarithmic corrections about the minimum $v_\Phi$.  This involves fixing the renormalisation scale to $\mu=v_\Phi$ and enforcing the condition $\lambda_{\rm eff}(v_\Phi)=0$, corresponding to the onset of radiative symmetry breaking \cite{Coleman:1973jx}. In this framework, the RG-improved one-loop potential is expressed as \cite{Iso:2017uuu,Marzo:2018nov,Ellis:2019oqb,Ellis:2020nnr} 
\begin{equation}
    V_{\rm eff}(\phi,T=0)\simeq \frac{\beta_{\lambda_{\rm eff}}}{4} \phi^4 \left(  \log \frac{\phi}{v_\Phi}-\frac{1}{4}\right), 
\end{equation}
where the running scalar quartic coupling is approximated by its $\beta$ function. The potential represents the leading-logarithmic resummation of the full RG-improved effective potential, valid near the vicinity of its minimum $\phi\sim v_\Phi$, at which the scalar quartic flips sign, resulting in a loop-suppressed mass of the scalar, $m_\Phi^2=\beta_{{\lambda_{\rm eff}}}v_\Phi^2$ with 
\begin{equation}\label{quartic}
    \beta_{\lambda_{\rm eff}}=\frac{\partial \lambda_{\rm eff} }{\partial \log \mu}=\frac{1}{16\pi^2}\left[ 96 {g^\prime}^4-2\lambda_{R}^4-\lambda_{\rm DM}^4\right].
\end{equation}

\begin{table}[t]
\centering
\setlength{\tabcolsep}{8pt} 
\renewcommand{\arraystretch}{1.3} 
\begin{tabular}{c c c c c c c c c c}
\hline
Field & $q_{L i}$ & $u_{R i}$ & $d_{R i}$ & $\ell_{L i}$ & $e_{R i}$ & $H$ & {\color{blue}$\boldsymbol{\Phi}$} & {\color{blue}$\boldsymbol{N_i}$} & {\color{red}$\boldsymbol{S}$} \\ \hline
$U(1)_{B-L}$ charges & $+\tfrac{1}{3}$ & $+\tfrac{1}{3}$ & $+\tfrac{1}{3}$ & $-1$ & $-1$ & $0$ & {\color{blue}$+2$} & {\color{blue}$-1$} & {\color{red}$+2$} \\ \hline
\end{tabular}
\caption{\it $U(1)_{B-L}$ charges of the fields in the minimal and extended $B-L$ models. The fields that extend the SM are shown in bold, with blue for the states in the minimal model and red for the extra state in the extended model.}
\label{table:BLcharges}
\end{table}

In the hot expanding universe ($T\neq0$), however, the effective potential receives temperature-dependent corrections from the 1-loop thermal and daisy re-summed contributions
\begin{equation}
\begin{aligned}
   {} V_T(\phi,T)&=3\frac{T^4}{2\pi^2}J_B\left( \frac{4{g^\prime}^2\phi^2}{T^2}\right)+4 \frac{T^4}{2\pi^2}J_F\left( \frac{\lambda_R^2\phi^2}{2T^2}\right)+ 2\frac{T^4}{2\pi^2}J_F\left( \frac{\lambda_{\rm DM}^2\phi^2}{2T^2}\right)\\
   & -\frac{2{g^\prime}^3}{3\pi}T \left[ (\phi^2+T^2)^{3/2}-\phi^3\right],
    \end{aligned}
\end{equation}
where we only include contributions from the $B-L$ gauge boson and the right-handed neutrinos, as the conformal scalar, with a vanishing tree-level mass, can only contribute at higher orders.

The thermal integrals are given as
\begin{equation}
    J_{B/F}(y)=\pm \int_0^\infty dx x^2 \log\left[1 \mp\exp\left( -\sqrt{x^2+y}\right) \right].
\end{equation}
Thus, the total effective potential up to the 1-loop plus the daisy resummation is given by
\begin{equation}
    V_{\rm total}(\phi,T)=V_{\rm eff} (\phi,T=0)+V_T(\phi,T).
\end{equation}
This state-of-the-art formulation of FOPT for the minimal $(B-L)$ model has already been extensively studied \cite{Jinno:2016knw, Iso:2017uuu}, and in this work, we followed the same prescription with precise evolution of the FOPT dynamics using a combined machinery of \texttt{PhaseTracer} \cite{Athron:2020sbe,Athron:2024xrh} and \texttt{TransitionSolver} \cite{Athron:2022mmm}.  

The dynamics of phase transitions are well-understood, with its theoretical foundations outlined in numerous earlier studies. In a thermal bath, phase transitions are primarily driven by thermal fluctuations, and the decay rate given by \cite{Linde:1981zj, Coleman:1977py}
\begin{equation} \label{rate}
    \Gamma(T)\simeq T^4 \left( \frac{S_3(T)}{2\pi T}\right)^{3/2}\exp \left( -\frac{S_3(T)}{T}\right),
\end{equation}
where the 3D Euclidean bounce action
\begin{equation}
    S_3(T)=4\pi\int_0^\infty dr\: r^2\left[  \frac{1}{2}\left(\frac{d\phi}{dr}\right)^2+V_{\rm total}(\phi,T)\right]. 
\end{equation}
To obtain the action, one needs to numerically solve the bounce equation of motion,
\begin{equation}
\frac{d^2 \phi}{dr^2}+\frac{2}{r}\frac{d\phi}{dr}=\frac{dV_{\rm total}(\phi,T)}{d\phi}.
\end{equation}
In our numerical analysis, the bounce action is solved using a modified version of \texttt{CosmoTransitions} \cite{Wainwright:2011kj} that is interfaced with \texttt{TransitionSolver}.

Note that there are substantial uncertainties in computations of the effective and nucleation rates \cite{Croon:2020cgk,Athron:2022jyi,Ekstedt:2023sqc}, as well as issues related to the gauge dependence of the effective potential \cite{Jackiw:1974cv,Wainwright:2011qy}. 
More extensive resummations \cite{Kierkla:2023von, Farakos:1994kx,Braaten:1995cm,Kajantie:1995dw,Ekstedt:2022bff,Dine:1992vs,Boyd:1993tz,Curtin:2016urg,Curtin:2022ovx} and directly computing the pre-factor as a functional determinant \cite{Ekstedt:2023sqc} may have a significant impact, but, even including these, significant uncertainties in the perturbative computations will persist\footnote{At least until a unified setup is developed that implements methods addressing all these issues  (see Refs.\  \cite{Patel:2011th,Hirvonen:2021zej,Lofgren:2021ogg} for gauge independent methods), while also tackling uncertainties in the use of the effective potential and the nucleation rate.}. Here, we focus on showing that in principle this model has a mechanism for dual explanations of dark matter and the baryon asymmetry of the universe that may be testable in the future, and forgo tackling these fundamental questions related to the uncertainty and robustness of the calculations, which we leave to future work.  
       
As the Universe transitions from its symmetric state, vacuum bubbles of the broken phase begin to form.
The critical temperature, denoted as $T_c$, is defined as the temperature at which the true and false vacuum are degenerate. When the temperature drops below $T_c$, large thermal fluctuations over the barrier in the effective potential can nucleate bubbles of the true vacuum. The bubbles
nucleate at a rate $\Gamma(T)$ (Eq.\ \eqref{rate}), and, using this, the  fractional volume of the false vacuum in the Universe can be computed with
\begin{equation} \label{fvfrac}
    p_f(T)=\exp[-I(T)]=\exp\left[  -\frac{4\pi v_w^3}{3} \int_{T}^{T_c}dT^\prime \frac{\Gamma(T^\prime)}{{T^\prime}^4H(T^\prime)} \left( \int_T^{T^\prime} \frac{d\tilde{T}}{H(\tilde{T})}\right)^3 \right],
\end{equation}
where we have assumed a fixed bubble wall velocity $v_w$ since throughout all the computations we will take $v_w\rightarrow 1$, as is expected for strongly supercooled phase transitions. The Hubble parameter evolves as
\begin{equation}
    H^2(T)=\frac{\overline{\rho}_{\rm tot}(T)}{3M_{\rm Pl}^2}
          = \frac{1}{3M_{\rm Pl}^2} \left(  \rho_{\rm tot}(\phi_f, T) - \rho(v_{\Phi}, 0)  \right) ,
          \label{Eq:HubbSq}
\end{equation}
where $M_{\rm Pl}=2.44\times 10^{18}$ GeV is the reduced Planck mass and we have used the bar on $\overline{\rho}_{\rm tot}(T)$ to indicate that we use the energy density in the false vacuum normalised to the zero temperature ground state, $v_{\Phi}$. 

In the high temperature expansion of the effective potential the dominant term is field independent and varies like $T^4$, and these contributions from all relativistic states give a radiation contribution to the energy density,  $\rho_R(T)=(\pi^2g_*(T)/30) T^4$ with $g_*(T)$ being the total number of relativistic degrees of freedom, which remains nearly constant within the temperature range of our interest.  Although we use the full expression given in Eq.\ \eqref{Eq:HubbSq} and integrate numerically in Eq.\ \eqref{fvfrac} to obtain the false vacuum fraction,  it is common in analytical calculations to assume the bag model equation of state holds approximately, so that the energy density can be written as $\overline{\rho}_{\rm tot}(T) \simeq \rho_R(T) + \rho_V(T) \simeq  \rho_R(T) + \Delta V(T)$ where $\rho_V(T)$ should be approximately constant allowing it to be replaced with $\Delta V(T)=V_{\rm total}(0,T)-V(v_{\Phi},T)$ at the end.

Tracking the value of false vacuum fraction, $p_f(T)$, provides important information on the progress of the phase transition.  At the beginning of the phase transition $p_f(T_c)=1$, while $p_f(T)$ approaching zero marks the completion of the phase transition to a new stable (true) vacuum.

Furthermore, during the phase transition, as the vacuum bubbles grow and occupy a larger fraction of the Universe’s total volume, they become connected, creating an infinite cluster that hinders a return to the original symmetric state.  This phenomenon, referred to as percolation, is expected to occur when $p_f(T_p)=0.71$ \cite{doi:10.1063/1.1338506, LIN2018299, LI2020112815}, defining the percolation temperature $T_p$. Since this is by definition related to contact between bubbles, it is a good choice for the temperature scale to use when evaluating thermal parameters that enter the predictions for the SGWB \cite{Athron:2023xlk}. The percolation temperature is found by numerically solving for $p_f(T_p)=0.71$ (or  equivalently $I(T_p)=0.34$), using Eq.\ \eqref{fvfrac}. To confirm the occurrence of percolation, it is crucial to ensure that, accounting for the expansion of space, the volume of space in the false vacuum is actually decreasing near $T_p$.  We test this by imposing \cite{Turner:1992tz,Ellis:2018mja,Athron:2022mmm}
\begin{equation}\label{completion}
   H(T)\left( 3+T\frac{dI(T)}{dT}\right)_{T_p}<0. 
\end{equation}
As the FOPT concludes, the energy released into the surrounding plasma causes the Universe to reheat back to a higher temperature, denoted as $T_{\rm reh}$. This process is particularly significant in scenarios with strong supercooling, due to the considerable amount of latent heat that is released. As a result, right after percolation occurs, the heavy scalar field $\Phi$ begins to oscillate around the true vacuum\footnote{This can be almost instantaneous or prolonged depending on the decay rate $\Gamma_\Phi$ \cite{Ellis:2019oqb}.} and will eventually decay. The reheating temperature can be expressed as follows
\begin{equation}
    T_{\rm reh}=
    \begin{cases}
    \left(\frac{\Gamma_\Phi}{H(T_p)}\right)^{1/2} T_p\left[  1+\alpha(T_p)\right]^{1/4},\:{\rm for}\: \Gamma_\Phi<H(T_p),\\
    T_p\left[  1+\alpha(T_p)\right]^{1/4},\:{\rm for}\: \Gamma_\Phi>H(T_p),
    \end{cases}
\end{equation}
where the parameter $\alpha$ is the strength of the phase transition and can be expressed as,
\begin{equation}
    \alpha=\frac{4\left(\bar{\theta}_{\rm false}(T_p)-\bar{\theta}_{\rm true}(T_p)\right)}{3\xi_{\rm false}},
\end{equation}
where the pseudotrace $\bar{\theta}$ is given by \cite{Giese:2020rtr}
\begin{equation}
    \bar{\theta}_i(T)=\frac{1}{4}\left(  \rho_i(T)-\frac{p_i(T)}{c_{s,i}^2(T)}\right),
    \end{equation}
with pressure $p=-V$, the enthalpy $\xi=p+\rho$, and the speed of sound $c_s$ in phase $\phi_i$,
\begin{equation}
    c_{s,i}(T)=\sqrt{\frac{\partial_TV}{T\partial_T^2V}\bigg|_{\phi_i(T)}}.
\end{equation}
In the bag model, there is a constant speed of sound $c_{s,i}=1/ \sqrt{3}$ and it is common to neglect deviations from this, so that $\bar{\theta}$ is replaced with $\theta = 1/4(\rho - 3 p)$.  Furthermore, if the supercooling is strong enough that the percolation temperature is much smaller than the the vacuum expectation value ($T_p \ll v_\Phi$), we can approximate both $\rho \approx \Delta V$ and $\theta \approx \Delta V$ greatly simplifying things (see e.g.\ Ref.\ \cite{Athron:2023rfq} for an explicit comparison of this and many other approximations). While we avoid these approximations in our numerical treatment, we have used them in analytical checks. 

Another crucial parameter, $\beta/H(T_p)$, represents the duration of the phase transition. This is often defined as $\beta = - dS/dT$, which appears in the first-order term of the Taylor expansion of the bounce action, and if one truncates the action at this order, it gives an exponential nucleation rate $\Gamma(t)\sim \exp(\beta t)$.  However, this is only valid for fast transitions, while for slow (i.e.\ strongly supercooled) transitions, there is a minimum in the action (leading to a maximum in the nucleation rate) such that $\beta$ can be zero or even negative during the phase transition. Here we instead calculate $\beta$ from the mean bubble separation, which at percolation is approximately given by\footnote{Technically this relation holds at the temperature $T_e$ defined by $P_f(T_e) = 1/e$ \cite{Enqvist:1991xw}.}, 
\begin{equation}
    \frac{\beta}{H}\bigg|_{T_p}=(8\pi)^{1/3}\frac{v_w}{R_{\rm sep} H(T_p)},
\end{equation}
 where the mean bubble separation is computed from the bubble number density, $n_B(T)$~\cite{Athron:2023xlk}: 
\begin{align}
  n_B(T) &= T^3 \!\! \int_T^{T_c} \! dT' \frac{\Gamma(T') p_f(T')}{T'^4 H(T')} \label{eq:bubble_num_density},\\
  R_{\text{sep}}(T) &= (n_B(T))^{1/3} . \label{eq:Rsep} 
\end{align}

\section{Scalar Field dynamics \& impact of RHN-mass} \label{s3}
After the phase transition, the scalar $\Phi$ can decay via three competative decay channels $\Phi\rightarrow N_iN_i$, $\Phi\rightarrow hh$ and $\Phi\rightarrow f\bar{f}V$. To generate an EMD phase, we want the decays to occur slowly enough that the scalar field can oscillate around its minima for an extended period, creating a matter-like era. When the VEV of the scalar is extremely large, the mixing through the Higgs portal becomes much weaker. However, a large gauge coupling favoured by conformal models can significantly enhance the radiative three-body decay into a SM fermion pair and SM gauge boson \cite{Blasi:2020wpy}.  If this decay is dominant, then any imprint on the duration of the EMD and the GWs will be insensitive to the RHN masses. To avoid this, we will require the RHN Yukawa couplings to be large enough that the decay to RHNs is dominant, but not so large that they spoil the occurrence of EMD. As a result, the duration of the EMD is primarily determined by the masses of the RHNs. For the dominant decays into the RHNs, the decay width is approximated by,
\begin{equation}\label{decay1}
    \sum_{i=1,2}\Gamma(\Phi\rightarrow N_iN_i)\approx\frac{2\lambda_R^2}{32\pi} m_\Phi,
\end{equation} 
while for decays into the Higgs, the approximate decay width is,
\begin{equation}
    \Gamma(\Phi\rightarrow hh)\approx\frac{\lambda_{H\Phi}^2v_\Phi^2}{32\pi m_\Phi},
\end{equation}
and the one-loop decay to SM fermions and vector bosons may be estimated by \cite{Blasi:2020wpy},
\begin{equation}\label{decay2}
    \Gamma(\Phi\rightarrow f\bar{f}V)\approx \beta_{\lambda_{\rm eff}}{g^\prime}^4 \left(\frac{m_\Phi}{10^8\rm GeV} \right).
\end{equation}
We neglect decays into DM since $\lambda_{\rm DM}\ll\lambda_{R}$. 

The end of EMD can be estimated from the dominant decay channel via $\Gamma_\Phi=H(T_{\rm dec})$, giving,
\begin{equation}
    T_{\rm dec}=\frac{1.2}{g_*(T_{\rm dec})^{1/4}} \sqrt{M_{\rm Pl}\Gamma_\Phi}.
\end{equation}

During the phase transition, the bubbles of the true vacuum nucleate and expand due to the pressure difference $\Delta V(T=0)$. In realistic scenarios, however, various plasma effects may limit the wall’s acceleration. The total pressure acting on the wall includes both the driving pressure $\Delta V$ and the frictional pressure due to particle interactions and gauge radiation \cite{Bodeker:2009qy,Gouttenoire:2021kjv}. If the driving pressure exceeds the maximum friction that can be exerted by the plasma, the wall would enter a runaway regime where its Lorentz factor increases without bound until bubbles collide, which is typically satisfied when the $T_p<T_{\rm eq}$, where $T_{\rm eq}$ is the vacuum-radiation equality temperature and is given as
\begin{equation}
    T_{\rm eq}= \left( \frac{30 \Delta V(T=0)}{\pi^2 g_*(T_{\rm eq})}\right)^{1/4},
\end{equation}
with $g_*(T_{\rm eq})$ denoting the effective number of relativistic degrees of freedom. During such runaway acceleration, the Lorentz factor $\gamma$ of the bubble wall grows approximately linearly with the bubble radius $R$ \cite{Gouttenoire:2021kjv} 
\begin{equation}
    \gamma(R)=\frac{2R}{3R_0},
\end{equation}
where $R_0$ is the initial bubble radius at nucleation, and is given by \cite{Ellis:2019oqb}
\begin{equation}
    R_0\equiv \left[ \frac{3 E_{0,V}}{4\pi \Delta V(T=T_n)}\right]^{1/3},
\end{equation}
where $E_{0,V}\simeq S_3(\phi_{\rm bounce})/2$ corresponds to the potential energy contribution to the energy of the initial bubble. For $R$ we use the mean bubble radius \cite{Megevand:2016lpr, Cai:2017tmh}
\begin{equation}
    R_{\rm mean}(T)=\frac{v_w T^2 }{n_B(T)} \int_{T}^{T_c} dT^\prime \frac{\Gamma(T^\prime)p_f(T^\prime)}{{T^\prime}^4 H(T^{\prime})}\int_T^{T^\prime}\frac{d\tilde{T}}{H(\tilde{T})}.
\end{equation}
We take the time of percolation to be when bubbles collisions are occurring and define the bubble radius at the time of collision as $R_{\rm coll}\equiv R_{\rm mean}(T_p)$.\footnote{At the time of collision $R_{\rm coll}\simeq R_{\rm sep}/2$.} Substituting this into the expression for $\gamma(R)$ yields the maximum Lorentz factor achievable in the runaway regime
\begin{equation}
    \gamma_{\rm max}\simeq 2R_{\rm coll}/3R_0,
\end{equation}

 During the bubble dynamics, the pressure difference across the wall can be expressed as
\begin{equation}
    \Delta P=\Delta V-P_{\rm LO}-P_{\rm NLO},
\end{equation}
where $P_{\rm LO}$ is the leading-order (LO) pressure accounting for $1\rightarrow 1$ scattering \cite{Bodeker:2009qy}, while $P_{\rm NLO}$  is
 the next-to-leading-order (NLO) contribution associated with $1\rightarrow N$  splittings in the vicinity of
 the bubble wall \cite{Hoche:2020ysm,Gouttenoire:2021kjv,Bodeker:2017cim}. For the $B-L$ model, it can be written as
 \begin{equation}
 \begin{aligned}
  {}   P_{\rm LO} &=\sum_jg_j c_j  \left( \frac{\Delta m_{j}^2T_p^2}{24}\right), c_j=\begin{cases}1\: {\rm for\: bosons}\\
  \frac{1}{2}\: {\rm for \: fermions}
  \end{cases}\\
  & = P_{\rm LO}^{(Z^\prime)}+\sum_iP_{\rm LO}^{(N_i)},
     \end{aligned}
 \end{equation}
 where the sum runs over particle species that experience a change in mass $\Delta m_j$ across the wall. Here, $g_j$ represents the number of degrees of freedom, with $g_{Z^\prime}=3$ and $g_{N_i}=2$, assuming that no other species gain mass as they cross the wall, and that the SM Higgs stays at $h=0$ during the tunnelling direction. There has been some controversy in computations of NLO friction with Refs.\ \cite{Bodeker:2017cim, Gouttenoire:2021kjv} finding it scales like $\gamma$ and Ref.\ \cite{Hoche:2020ysm} finding a $\gamma^2$ scaling.  While Ref.\ \cite{Gouttenoire:2021kjv} argues that the discrepancy is connected to the Ward identity being satisfied in Ref.\ \cite{Hoche:2020ysm} in situations where it should be broken by the presence of the bubble wall, we will remain agnostic on this issue.  We will  focus solely on the dominant contribution from the $Z^\prime$ boson, and compare both debated possibilities of NLO friction, which can be expressed as follows
 \begin{equation}
     P_{\rm NLO}^{(2)}\sim \gamma^2 \mathcal{F}_{\rm eff}\frac{3\zeta_3(2\log2-1)}{32\pi^4}{g^\prime}^2 T_p^4,
 \end{equation}
 and 
 \begin{equation}
     P_{\rm NLO}^{(1)}\sim \gamma\mathcal{F}_{\rm eff}\frac{3\kappa\zeta_3}{8\pi^4}{g^\prime}^3 v_\Phi T_p^3 \log\left( \frac{v_\Phi}{T_p}\right),
 \end{equation}
 where $\zeta_3=1.20206$ and $\mathcal{F}_{\rm eff}$ accounts for the effective number of relativistic degrees of freedom emitting soft $Z^\prime$
 quanta at $T_p$. The factor $\kappa\approx4$ \cite{Gouttenoire:2021kjv} accounts for the two possible emission legs (before and after scattering) and the two transverse polarisations of the emitted gauge boson in the $1\rightarrow2$ splitting processes within the plasma.  In our model, the dominant contributions to $\mathcal{F}_{\rm eff}$ arise only from Standard Model fermions and two RHNs, yielding
 \begin{equation}
 \begin{aligned}
 {}  \mathcal{F}_{\rm eff} &= \sum_{f} N_c^{(f)} g_{\rm spin}^{(f)} Q_{B-L}^2(f), \:{\rm with}\: \begin{cases}
 N_c^{(f)}=3,g_{\rm spin}^{(f)}=2,\:{\rm for\:quarks}\\
 N_c^{(f)}=1,g_{\rm spin}^{(f)}=2,\:{\rm for\:charged \:leptons \:and \: LH\:neutrinos}\\
 N_c^{(f)}=1,g_{\rm spin}^{(f)}=1,\:{\rm for\:RH\: neutrinos}\\
 \end{cases}\\
 &\simeq 28,
    \end{aligned}
 \end{equation}
where $N_c^{(f)}$ and $g_{\rm spin}^{(f)}$ denote the color and spin multiplicities, respectively, while $Q_{B-L}(f)$ is the $U(1)_{B-L}$ charge of the fermion $f$ (see Table \ref{table:BLcharges}).
\\
 If the wall expands further, the NLO terms become important, and finally, the wall reaches a stationary state where any pressure difference vanishes, leading to the terminal Lorentz factor
\begin{equation}
    \gamma_{\rm LL}=\left(\frac{\Delta V-P_{\rm LO}}{P_{\rm NLO}^{(c)}}\right)^{1/c},
\end{equation}
where $c=1,2 $ depending on the scaling of the $1\rightarrow N$ NLO pressure, $P_{\rm NLO}\propto \gamma^c$. 

Finally, the effective Lorentz factor at the moment of bubble collision is therefore set by the smaller of these two limiting regimes,
\begin{equation}
    \gamma_{\rm coll} \simeq \min\left[ \gamma_{\rm max}, \, \gamma_{\rm LL} \right],
\end{equation}
  which determines the energy available for subsequent processes, such as GW generation or nonthermal particle production. We will discuss these processes in Sections \ref{s5}, \ref{s4} respectively.


At the time of bubble percolation, the latent heat of the universe has been converted into two fluids, the radiation-like energy density $\rho_{\rm plasma}$ of the ultra-relativistic shock waves generated by the work of the friction pressure, and the (initially) radiation-like energy density of the scalar field gradient $\rho_\Phi$.  If the scalar field is long-lived enough, then the scalar field energy density starts redshifting like matter, with an energy density after percolation given by
\begin{equation}
    \rho_{\Phi}(t)=\kappa_{\rm coll} \Delta  V \left( \frac{a(t_p)}{a(t)}\right)^3,
\end{equation}
which translates to the temperature at which the energy density for the scalar field dominates, 
\begin{equation}\label{Tdomeq}
    T_{\rm dom}\sim \kappa_{\rm coll} T_{\rm eq},
\end{equation}
where \cite{Ellis:2019oqb,Ellis:2020nnr}
\begin{equation}
    \kappa_{\rm coll}= \frac{2}{3}\left(  1-\frac{\alpha_\infty}{\alpha}\right)\frac{\gamma_{\rm coll}}{\gamma_{\rm max}},
    \label{Eq:kappaColl}
\end{equation}
where $\alpha_\infty =P_{\rm LO}/\rho_R(T_p)$ \cite{Ellis:2019oqb}. Note also here we use $\Delta V$ for the energy liberated from the vacuum because during the period of vacuum domination, the $T$ dependent parts can be neglected. 
A transition from matter to radiation domination introduces entropy into the universe. This increase in entropy dilutes the abundance of any relics that were present in the universe before the scalar decay, given roughly by the dilution factor
\begin{equation}\label{entropy}
    \Delta=\frac{S_{\rm dec}}{S_{\rm dom}}\simeq 1+\frac{T_{\rm dom}}{T_{\rm dec}}.
\end{equation}
We will discuss this effect on the GW spectrum in the following Sec.\ \ref{s5}. 

\subsection{Singlet extension: suppressed $\Phi\rightarrow f\bar{f}V$ channel}\label{singlet_ext}

If the minimal model is extended by introducing an additional complex scalar singlet $S$ \cite{Bian:2019szo,Huang:2022vkf}, carrying the same $B-L$ charge as $\Phi$ (see Table \ref{table:BLcharges}), yet remaining inert—meaning it does not acquire a VEV and therefore does not contribute to the RHN masses. The renormalised quartic coupling receives a modified beta function,
\begin{equation}
     \beta_{\lambda_{\rm eff}}^{\rm mod}=\frac{\partial \lambda_{\rm eff} }{\partial \log \mu}=\frac{1}{16\pi^2}\left[\lambda_{\Phi S}^2 +96 {g^\prime}^4-2\lambda_{R}^4-\lambda_{\rm DM}^4\right].
\end{equation}
This contribution can keep the running of the quartic at the same order as in Eq. \eqref{quartic}, even for slightly smaller values of $g^\prime$. Consequently, the gauge-mediated decay channel $\Phi\rightarrow f\bar{f}V$ becomes suppressed, making the dominant decay mode $\Phi \rightarrow N_i{N_i}$ as long as the kinematic condition $m_\Phi<2m_{Z^\prime}$ is satisfied. This requirement implies the following approximate upper bound on the scalar mixing coupling
\begin{equation}
    \lambda_{\Phi S}^2< 64\pi^2{g^\prime}^2-96{g^\prime}^4.
    \end{equation}
The modified LO pressure in this case can be written as,
 \begin{equation}
 \begin{aligned}
  {} &   P_{\rm LO}=P_{\rm LO}^{(Z^\prime)}+\sum_iP_{\rm LO}^{(N_i)}+P_{\rm LO}^{(S)},
     \end{aligned}
 \end{equation}
where $g_S=2$ for a complex scalar. The NLO pressure terms remain roughly the same since the scalar mass is large, and any contributions from such emission will be Boltzmann suppressed. 
 
\section{Thermal Leptogenesis \&  non-thermal DM from scalar decay}\label{s4}

After the electroweak phase transition, when the SM Higgs field acquires its vacuum expectation value
$\langle H\rangle=v_h$, the first and third terms in Eq.~\eqref{Leff} generate the light active neutrino masses through the standard type-I seesaw mechanism, $m_{\nu i}\sim {Y_i^\nu}^2v_h^2/M_N$ \cite{seesaw1,seesaw2,seesaw3,seesaw4}. At much higher temperatures, $T\gg T_{\rm EW}$, the out-of-equilibrium and CP asymmetric decay of heavy RHNs into lepton doublets and Higgs fields, produces a net lepton asymmetry. The asymmetry generated by the $i$-th RHN freezes out when $T\sim M_i$, setting the characteristic temperature scale for thermal leptogenesis. 

In the conventional high-scale leptogenesis picture, three distinct flavour regimes are identified, depending on which charged-lepton Yukawa interactions (with an interaction strength $\Gamma_\alpha\sim 5\times 10^{-3}Y_{L\alpha}T$, where $Y_{L\alpha}$ is the charged lepton Yukawa coupling) are in equilibrium \cite{fllep1,fllep2,fllep3}. At $T\gtrsim 10^{12}$ GeV, all charged-lepton Yukawas are effectively out of equilibrium, and leptogenesis proceeds in the one-flavour (1FL) or ``vanilla'' regime, where the produced lepton states remain coherent superpositions $| L_i \rangle=\mathcal{A}_{i\alpha}| L_\alpha \rangle$ with amplitudes $\mathcal{A}_{i\alpha}$. In this regime, the final baryon asymmetry is insensitive to the low-energy neutrino phenomenology and CP phases.

At lower temperatures, flavour effects become important. When $T\lesssim 10^{12}$ GeV, the $\tau$-Yukawa interactions enter equilibrium, breaking coherence and splitting the state into $\tau$ and $(e+\mu)$ components, defining the two-flavour (2FL) regime. When $T\lesssim 10^9$ GeV, the $\mu$-Yukawa interactions also equilibrate, leading to the three-flavour (3FL) regime. In these flavoured regimes, the efficiency of leptogenesis can depend on low-energy neutrino parameters, and, as we show, different flavour regimes can leave distinct imprints on the SGWB spectrum through their connection to the RHN mass and the resulting matter-dominated dynamics.

After the supercooled first-order phase transition, a fraction $\kappa_{\rm coll}$ of the released latent heat $\Delta V$ is converted into the scalar field energy density, so,
\begin{equation}
    \rho_\Phi(T_p)=\kappa_{\rm coll}\Delta V,
\end{equation}
and the radiation energy density released in the plasma at percolation 
\begin{equation}
    \rho_{\rm plasma}(T_p)=(1-\kappa_{\rm coll})\Delta V.
\end{equation}
The two energy components evolve as the Universe expands,
\begin{equation}\label{e_density}
    \begin{aligned}
        {}  \rho_\Phi(t)&=\kappa_{\rm coll}\Delta V \left(  \frac{a_p}{a(t)}\right)^3,\\
          \rho_{\rm plasma}(t)&=(1-\kappa_{\rm coll})\Delta V \left(  \frac{a_p}{a(t)}\right)^4.
    \end{aligned}
\end{equation}
At the time of scalar field domination at $T=T_{\rm dom}$, $\rho_\Phi(T_{\rm dom})=\rho_{\rm plasma}(T_{\rm dom})$, which gives
\begin{equation}\label{red_shift}
    \frac{a_p}{a_{\rm dom}}=\frac{\kappa_{\rm coll}}{1-\kappa_{\rm coll}}.
\end{equation}
Using Eq.\ \eqref{e_density} and \eqref{red_shift}, one can get an estimate of the scalar field energy density when it dominates the energy density
\begin{equation}
    \rho_\Phi(T_{\rm dom})=\frac{\kappa_{\rm coll}^4}{(1-\kappa_{\rm coll})^3}\Delta V.
\end{equation}
When the scalar decays, a fraction of its energy is transferred to a pair of RHNs, one of which can serve as DM. The RHN DM number density, $n_{\rm DM}$, at decay is, 
\begin{equation}
\begin{aligned}
{} n_{\rm DM}(T_{\rm dec})&= {\rm BR}(\Phi\rightarrow N_{\rm DM}  N_{\rm DM}) \, n_\Phi(T_{\rm dec})\\
&  =\frac{\lambda_{\rm DM}^2}{2\lambda_{R}^2+\lambda_{\rm DM}^2}\frac{\rho_{\Phi}(T_{\rm dom})}{m_\Phi}\left( \frac{T_{\rm dec}}{T_{\rm dom}}\right)^3,
\end{aligned}
\end{equation}
where $n_\Phi = \rho_{\Phi} / m_\Phi$ is the number density of $\Phi$. 
The present-day DM relic abundance is then given by,
\begin{equation}
    \Omega_{\rm DM}h^2=\frac{m_{\rm DM}n_{\rm DM}(T_{\rm dec})}{s(T_{\rm dec})}\frac{s_0}{\rho_c/h^2}.
\end{equation}
However, such a non-thermal RHN DM can have a large velocity at the standard matter radiation equality unless $T_{\rm dec}$ is sufficiently high, such that its velocity $v_{\rm DM}(t)$ is red-shifted \cite{Takahashi:2007tz,BhupalDev:2013oiy,Hisano:2000dz}. The large velocity might result in a large co-moving free-streaming length (horizon) $\lambda_{\rm FS}$ which is constrained by the Ly-$\alpha$ cloud as $\lambda_{\rm FS}\lesssim 1$ Mpc \cite{BhupalDev:2013oiy}. This translates into a lower bound on the DM mass (see appendix \ref{APP-A} for a detailed derivation),
\begin{equation}
     M_{\rm DM}\gtrsim 1015.25\left(\frac{g_{*s}^{\rm eq}}{g_{*s}^{\rm dec}} \right)^{1/3}\left( \frac{T_{\rm eq}}{T_{\rm dec}}\right)m_\Phi.
\end{equation}
\section{Modified GW spectrum from RHN-mass dependent matter domination}\label{s5}

During the strongly supercooled PT, the large latent heat $\Delta V$ is converted into scalar field gradients. This drives bubble walls to ultra-relativistic Lorentz factors.  The pressure driving the acceleration is opposed by friction that grows with the velocity\footnote{As discussed earlier, it grows like $\gamma$ or possibly $\gamma^2$.} and one also expects some of the energy to transfer to the surrounding plasma.  In fact, typically in first-order cosmological PTs most of the energy is transferred to the plasma, but in scenarios with extreme supercooling boosting $\alpha$ to very large values, it is possible for most of the energy to reside in the scalar field gradients (i.e.\ in the bubble walls) \cite{Ellis:2019oqb}.  Gravitational waves can be sourced from both the energy stored in the bubble walls and the energy in the plasma and may potentially give rise to signals that could be detected in future experiments.  

The relative size of the GW signals generated from different sources depends on the fraction of energy that goes into each one, which can be quantified through efficiency coefficients like $\kappa_{\text{coll}}$ (see Eq. \ref{Eq:kappaColl}) for the bubble collisions, with $1-\kappa_{\text{coll}}$ available for plasma sources\footnote{Some energy will also go to heat.} (sound waves and turbulence).  In our scenarios, we do have ultra-relativistic Lorentz factors, but the $\alpha$ is not as large as the values where Ref.\ \cite{Ellis:2019oqb} found bubble collisions to be the dominant source. Thus, we expect plasma sources to dominate, but our scenarios with very large $\alpha$ lie far beyond the validity of 3D hydrodynamical simulations for sound waves \cite{Hindmarsh:2017gnf}, and while many recent advances have been made with the sound shell model \cite{Hindmarsh:2016lnk,Hindmarsh:2019phv} it is unclear how well it models our scenarios.  Furthermore, turbulence has been found to contribute significantly to observability \cite{Ellis:2019oqb}, but it is the least well modelled/understood of these sources (see e.g.\ Ref.\ \cite{Athron:2023xlk} and references therein). Other approaches have also been considered e.g.\  \cite{Konstandin:2017sat, Lewicki:2020azd, Lewicki:2022pdb}
 but currently, there remains a substantial uncertainty in predicting the gravitational wave spectra for given thermal parameters, in addition to the potentially even larger uncertainties from the computation of thermal parameters from the effective potential \cite{Athron:2023rfq}, and the effective potential itself \cite{Croon:2020cgk,Athron:2022jyi}.   
 
 Here we follow Refs.\ \cite{Gouttenoire:2023bqy,Gouttenoire:2023pxh} and use the bulk flow model \cite{Jinno:2017fby,Konstandin:2017sat,Cutting:2020nla}, with some modifications we will discuss below, to predict the gravitational waves spectrum (see Ref.\ \cite{Gouttenoire:2023bqy} for a justification of this choice).  Assuming an adiabatic radiation-dominated universe after percolation, the spectrum today can be described by \cite{Konstandin:2017sat},
\begin{equation}
    \Omega_{\rm GW}h^2=\frac{h^2}{\rho_c}\frac{d\rho_{\rm GW}}{d\ln f}\simeq \frac{10^{-6}}{(g_*/100)^{1/3}} \left(\frac{\beta}{H_p} \right)^{-2}\left( \frac{\alpha}{1+\alpha}\right)^2 S(f) S_H(f),
\end{equation}
where the spectral shape function is given by
\begin{equation}
    S(f)=\frac{3(f/f_{\rm peak})^{0.9}}{2.1+0.9(f/f_{\rm peak})^3},
\end{equation}
which peaks at
\begin{equation}
    f_{\rm peak}=0.8\left( \frac{a_p}{a_0}\right)\left( \frac{\beta}{2\pi}\right).
\end{equation}
The redshift factor $\frac{a_p}{a_0}$ relates the scale factor at percolation to today
\begin{equation}
    \frac{a_p}{a_0}=1.65\times 10^{-5} \:{\rm Hz} \left(\frac{T_{\rm eq}}{100\:\rm GeV}\right)\left(\frac{g_*(T_{\rm eq})}{100}\right)^{1/6}H_p^{-1}.
\end{equation}
The correction factor $S_H(f)$ ensures the correct causal scaling ($\Omega_{\rm GW}\propto f^3$) for frequencies corresponding to modes that were super-horizon at the time of emission
\begin{equation}
    S_H(f)=\frac{(f/f_H)^{2.1}}{1+(f/f_H)^{2.1}}, \: {\rm with}\: f_H=c_*\left(\frac{a_p}{a_0}\right)\left(\frac{H_p}{2\pi}\right),
\end{equation}
where $c_*=\mathcal{O}(1)$ is a constant.

However, a critical modification arises in models if the scalar field responsible for the phase transition is sufficiently long-lived \cite{Ellis:2020nnr,Gouttenoire:2023pxh}. The energy density of GWs, which scales as $\rho_{\text{GW}} \propto a^{-4}$, is diluted relative to the entropy density $s \propto a^{-3}$ by the factor $\Delta$. The GW abundance 
is therefore suppressed by $\Delta^{4/3}$ and the wavelengths of the GWs are stretched by the entropy injection, causing a net red-shift of the entire spectrum by a factor of $\Delta^{1/3}$. The low-frequency, causal tail of the spectrum is altered. Modes with wavelengths that entered the cosmological horizon during the MD era inherit a different scaling. Instead of the standard $\Omega_{\text{GW}} \propto f^3$ tail \cite{Durrer:2003ja,Caprini:2009fx,Cai:2019cdl} for modes entering during radiation domination, the slope becomes $\Omega_{\text{GW}} \propto f^1$ \cite{Barenboim:2016mjm,Ellis:2020nnr}. This is encoded in an additional correction factor $S_M(f)$ applied to the spectrum for frequencies $f_{\text{dec}} < f < f_{\text{dom}}$, where
\begin{equation}
    f_{\rm dom} =\left( \frac{a_{\rm dom}}{a_0}\right)\frac{H_{\rm dom}}{2\pi},\: f_{\rm dec}=\left( \frac{a_{\rm dec}}{a_0}\right)\frac{H_{\rm dec}}{2\pi}.
\end{equation}
Thus, the modified GW spectrum today, accounting for the EMD era, is given by
\begin{equation}
    \Omega_{\rm GW}^{\rm EMD}h^2 \simeq \frac{1}{\Delta^{4/3}} \frac{10^{-6}}{(g_*/100)^{1/3}} \left( \frac{\beta}{H_p}\right)^{-2} \left( \frac{\alpha}{1+\alpha}\right)^2 S(\Delta^{1/3}f) S_H(\Delta^{1/3}f) S_M(f).
\end{equation}

  \begin{figure}
\centering
\includegraphics[scale=.68]{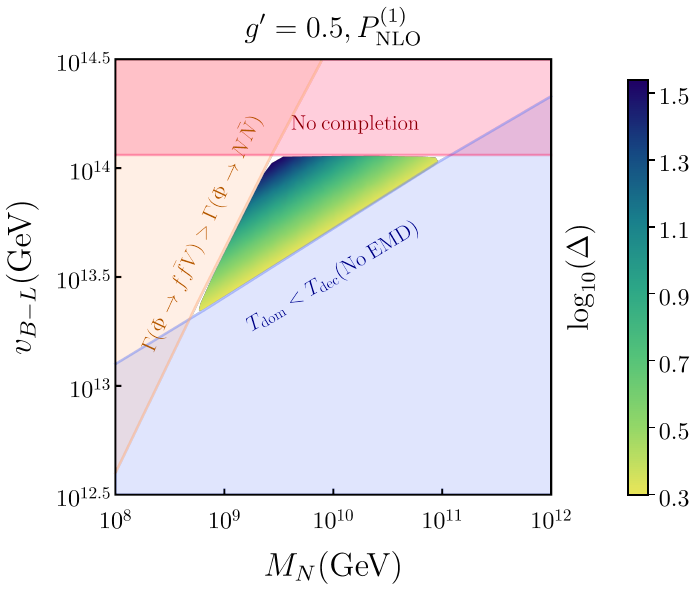}
\includegraphics[scale=.68]{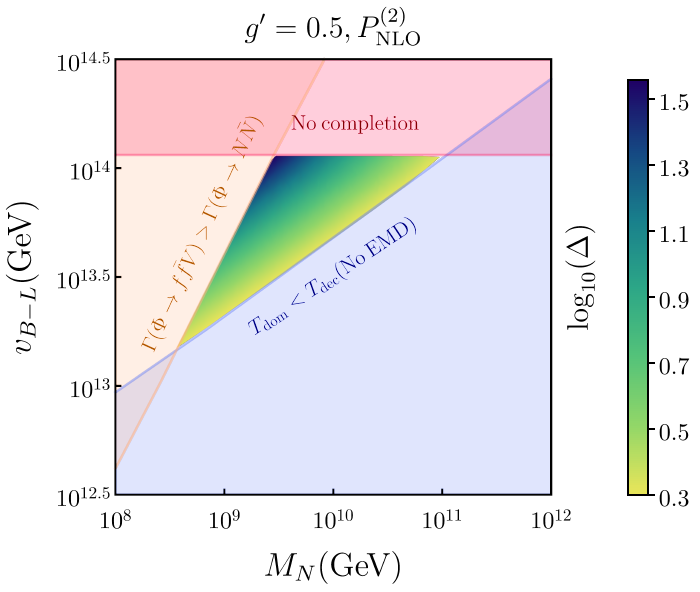}
\includegraphics[scale=.68]{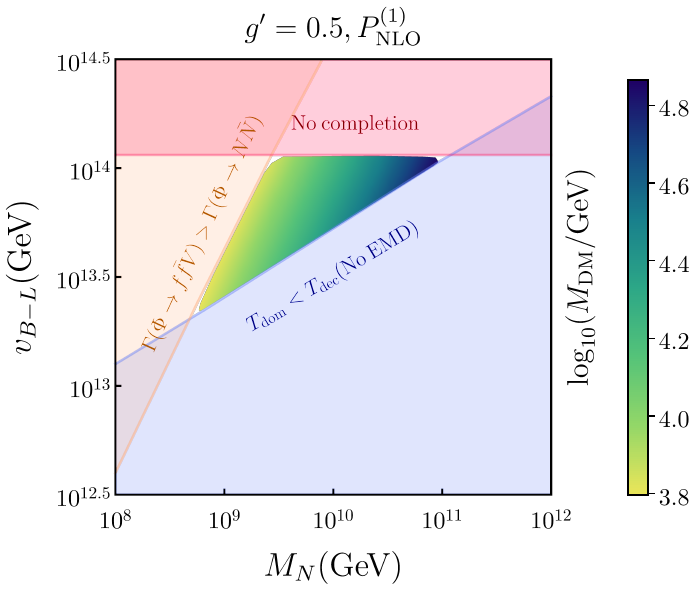}
\includegraphics[scale=.68]{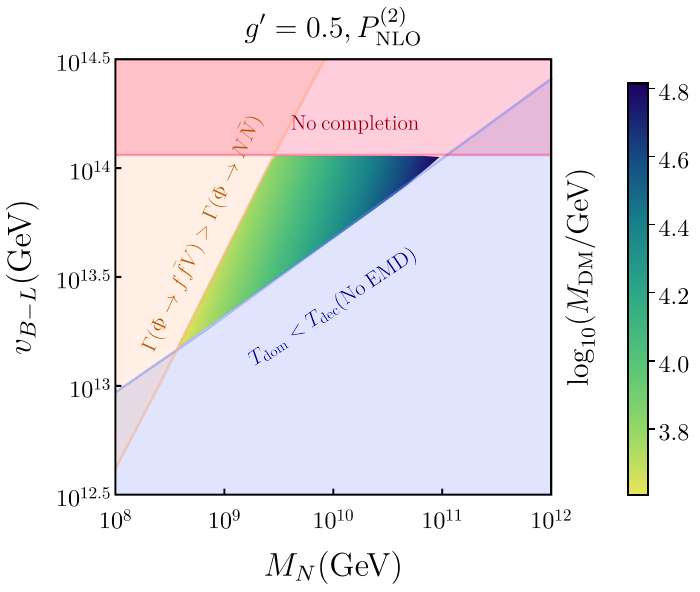}
 	\caption{\textit{Top-left:} The allowed parameter space for RHN-mass controlled matter domination, assuming $P_{\rm NLO}^{(1)}$ friction with $g^\prime=0.5$, is indicated where the green shaded region represents the amount of entropy production after the end of early matter domination. \textit{Top-right:} The same parameter space as left, but for $P_{\rm NLO}^{(2)}$ friction. \textit{Bottom-left:} The allowed parameter space for RHN-mass controlled matter domination, assuming $P_{\rm NLO}^{(1)}$ friction with $g^\prime=0.5$, is indicated where the green shaded region represents the RHN dark matter mass required to achieve the correct relic density, $\Omega_{\rm DM}h^2 \simeq 0.12$. \textit{Bottom-right:}  The same parameter space as left, but for $P_{\rm NLO}^{(2)}$ friction.}\label{fig1}
 \end{figure}

\section{Numerical Results}\label{s6}

In this section, we explore the parameter space where thermal leptogenesis and non-thermal RHN dark matter can coexist consistently within the conformal $U(1)_{B-L}$ framework. Our primary goal is to identify regions in which the post-transition cosmology leaves a distinctive RHN-mass-dependent imprint on the primordial GW spectrum. This feature arises because the same parameters that govern leptogenesis—the RHN mass spectrum and Yukawa couplings—also control the scalar decay dynamics that determine the reheating history and entropy production after the phase transition. To make this connection transparent, we first focus on a benchmark gauge coupling $g^\prime =0.5$, which is close to the lowest possible value in the minimal model. At this point, the transition is strongly supercooled, and the resulting dynamics are representative of the model’s most predictive regime. Larger gauge couplings increase the overall viable region but simultaneously push the RHN masses to higher values, rendering the scenario less phenomenologically accessible. We therefore next consider the impact of reducing $g^\prime$, which necessitates a singlet extension of the minimal model. This modification simultaneously alters the strength of the phase transition and the branching ratios of the scalar decay channels in Eqs.~\eqref{decay1}–\eqref{decay2}. This variation directly affects the reheating efficiency and the duration of the ensuing EMD, thereby modulating the shape of the GW spectrum.

In Fig.\ \ref{fig1}, we present the allowed parameter space on $M_N-v_{B-L}$\footnote{For notational consistency with the plotted variables, we employ $v_{B-L}$ in this section, where $v_{B-L} \equiv v_\Phi$.} plane considering both possibilities of NLO friction terms, i.e.\ $P_{\rm NLO}^{(1)}$ (left) or $P_{\rm NLO}^{(2)}$ (right). The red shaded region fails the completion criterion (Eq.\ \eqref{completion}) and the blue region does not have a period of EMD because the scalar decays too quickly to dominate the energy density, making it irrelevant for our scenario. Similarly, the orange region is where the three-body decay $\Phi\rightarrow f \bar{f}V$ is more dominant than the RH neutrino pair production, in violation of our assumptions and corresponding to scenarios where the RHN masses would not influence the duration of the EMD or the GWs spectral shapes. While the upper panels show the resulting entropy injection depending on the strength of matter domination, the lower panels show the non-thermal RHN DM mass required to reproduce the correct observed relic abundance. 
   \begin{figure}
\centering
\includegraphics[scale=.68]{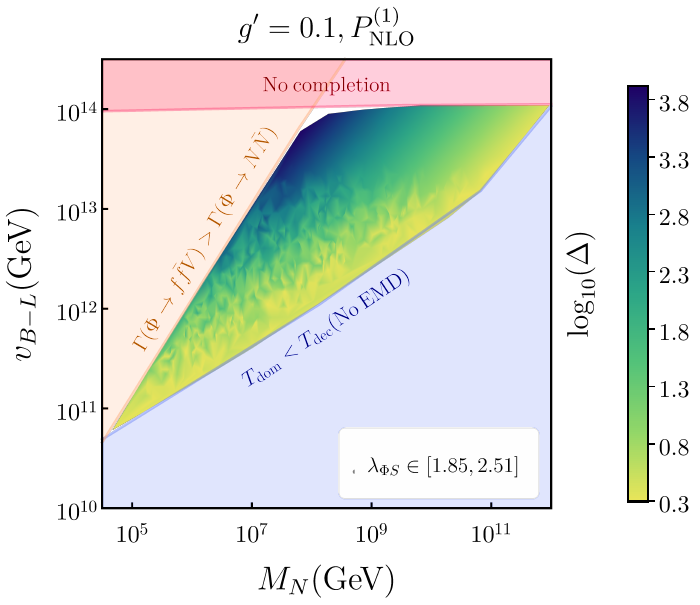}
\includegraphics[scale=.68]{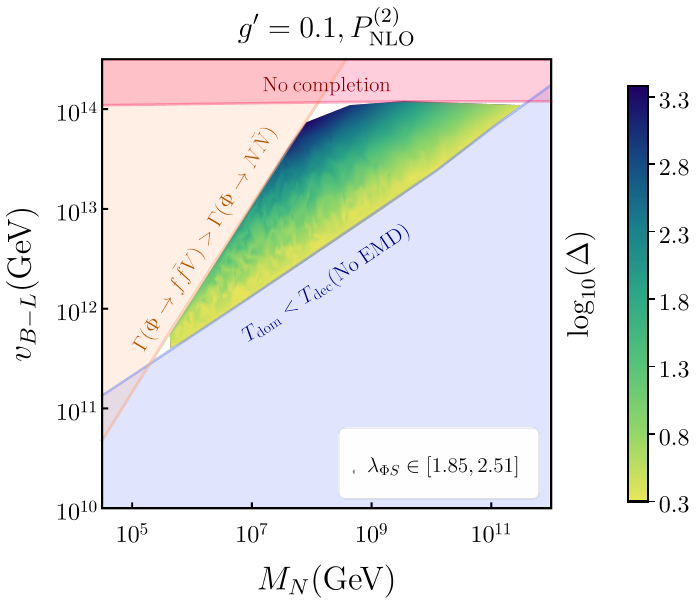}
\includegraphics[scale=.68]{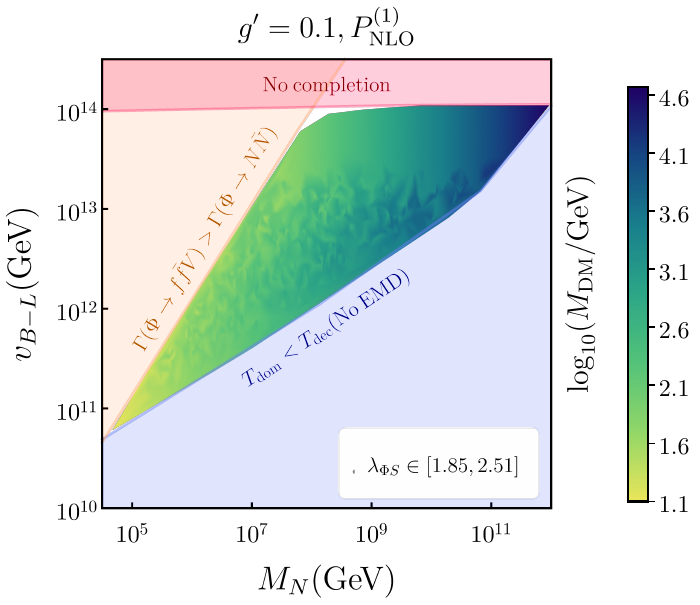}
\includegraphics[scale=.68]{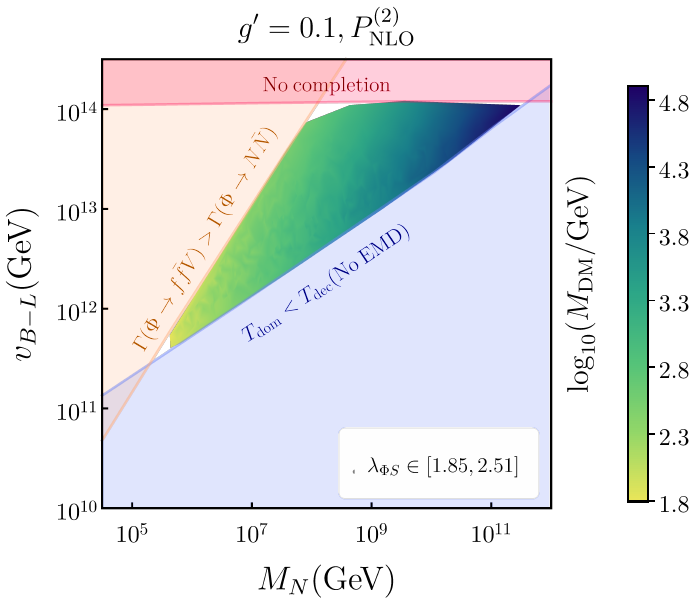}
 	\caption{\textit{Top-left:} The allowed parameter space for RHN-mass controlled matter domination with the singlet extension, assuming $P_{\rm NLO}^{(1)}$ friction with $g^\prime=0.1$ and $\lambda_{\Phi S}\in [1.85,2.51]$, is indicated where the green shaded region represents the amount of entropy production after the end of early matter domination. \textit{Top-right:} The same parameter space as left, but for $P_{\rm NLO}^{(2)}$ friction. \textit{Bottom-left:}The allowed parameter space for RHN-mass controlled matter domination with the singlet extension, assuming $P_{\rm NLO}^{(1)}$ friction with $g^\prime=0.1$ and $\lambda_{\Phi S}\in [1.85,2.51]$, is indicated where the green shaded region represents the RHN dark matter mass required to achieve the correct relic density, $\Omega_{\rm DM}h^2 \simeq 0.12$. \textit{Bottom-right:}  The same parameter space as left, but for $P_{\rm NLO}^{(2)}$ friction.}\label{fig2}
 \end{figure}
Two physical trends are immediately visible from the plots, and they are worth stressing explicitly. First, at fixed $g^\prime$, increasing $v_{B-L}$ pushes the reheating temperature upward, which raises the characteristic temperature $T_{\rm dom}$ (see Eq.\ \eqref{Tdomeq}) at which the scalar field's energy density would begin to dominate, while decreasing $M_N$ suppresses the scalar decay width and so lowers the decay temperature $T_{\rm dec}$. The net result is that entropy production grows (according to Eq.\ \eqref{entropy}) toward higher $v_{B-L}$ and lower $M_N$. Second, the apparent insensitivity of the required RHN DM mass to $T_{\rm dec}$ or $M_N$ in large swathes of parameter space is a consequence of an approximate cancellation between redshift factors and the entropy scaling evaluated at $T_{\rm dec}$. However, increasing $M_N$ reduces the scalar branching fraction into DM, so a larger DM mass is required to satisfy the DM relic abundance.  

 When the model is extended with an additional singlet, the renormalised scalar-quartic is mostly dominated by the $\lambda_{\Phi S}$ for smaller $g^\prime$. Although this keeps the overall magnitude of the quartic effectively the same as the minimal scenario, the smaller $g^\prime$ suppresses the $\Phi\rightarrow f\bar{f}V$ channel relative to the RHN-dominated channel (see Sec.\ \ref{singlet_ext} for details). This suppression of the three-body decay width, together with the condition $m_\Phi<m_{Z^\prime}$, significantly enlarges the viable parameter space, as shown in Fig.\ \ref{fig2}. However, we should note that models with a very small $g^\prime$, or those in which local symmetry is replaced by global symmetry, will not have a significant contribution from the $\Phi \rightarrow f\bar{f}V$ channel, or this channel may be entirely absent. Consequently, the scalar era can be fully controlled by the RHN mass. This aspect is phenomenologically interesting, especially considering the detectability of GW in such scenarios during upcoming experiments, which we plan to address in a follow-up publication.
    \begin{figure}
\centering
\includegraphics[scale=.72]{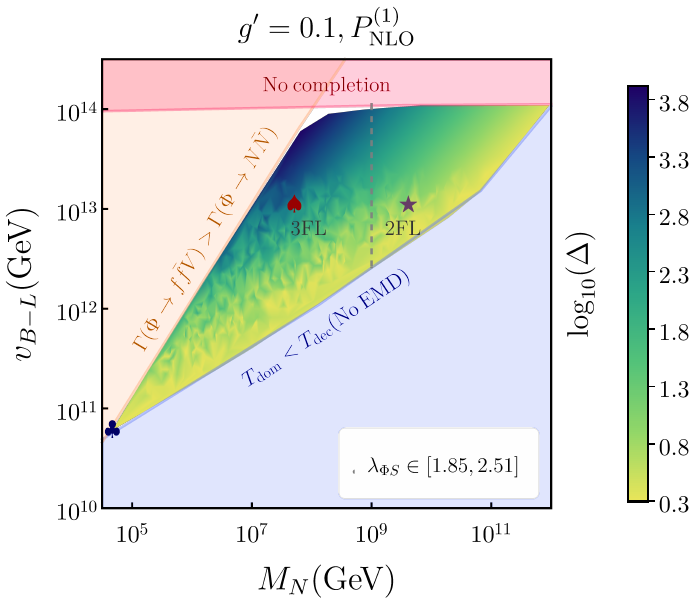}
\includegraphics[scale=.67]{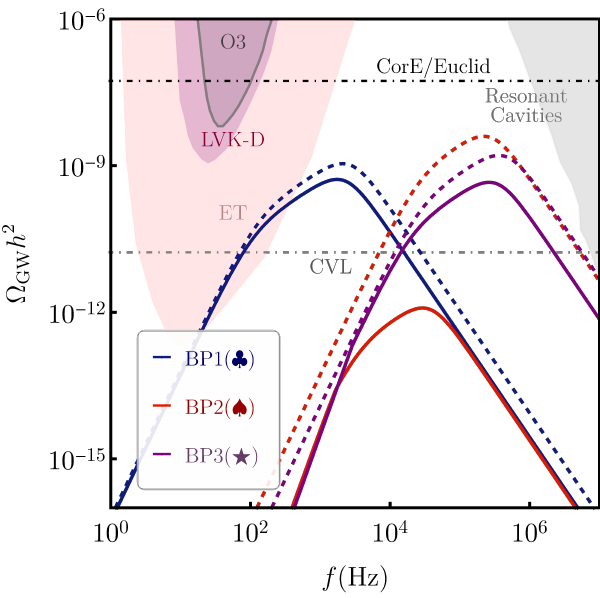}
 	\caption{\textit{Left:} The benchmarks BP1 ($\clubsuit$), BP2 ($\spadesuit$), BP3($\bigstar$) correspond to different flavour regimes of leptogenesis, assuming $P_{\rm NLO}^{(1)}$ friction with $g^\prime=0.1$ and $\lambda_{\Phi S}\in [1.85,2.51]$, where the green shaded region represents the amount of entropy production after the end of early matter domination. \textit{Right:} The GW spectrum corresponding to the benchmarks on the left is shown as solid lines (with modified reheating) and dashed lines (without modified reheating).}\label{fig3}
 \end{figure}
 
Remarkably, as can be seen from the plots, the different ranges for $M_N$ correspond to the different flavour regimes of leptogenesis. We find that in the minimal model, the viable parameter space shrinks considerably for $g^\prime\lesssim 0.5$, confining $M_N$ to values above approximately $10^9$ GeV. Increasing $g^\prime$ enlarges the allowed region but simultaneously raises the lower bound on $M_N$. Thus, we can conclude that in the minimal scenario, RHN-mass–dependent matter domination cannot be realised for $M_N\lesssim 10^9$ GeV, implying that such signatures are absent in the 3FL. Nonetheless, as shown in Fig.\ \ref{fig1}, for $g^\prime=0.5$ substantial portion of the 2FL regime is testable. For a fixed VEV, the entropy production increases as $M_N$ decreases, a feature that becomes crucial in distinguishing the different flavour regimes defined by the RHN mass.

\begin{figure}
\centering
\includegraphics[scale=.75]{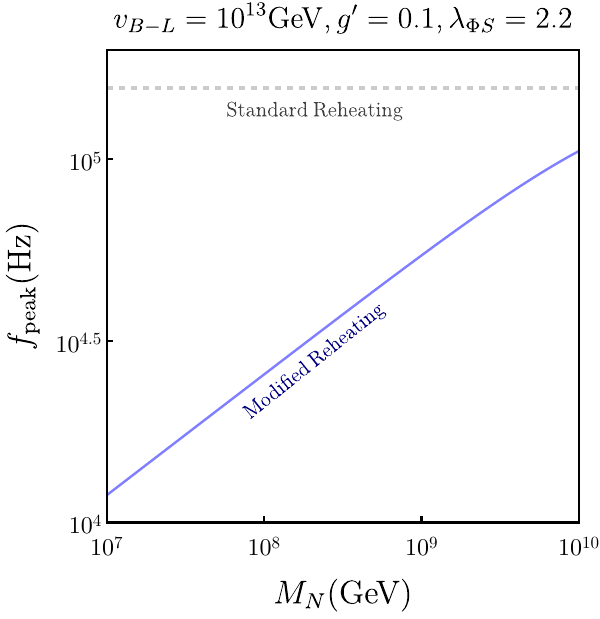}
\includegraphics[scale=.75]{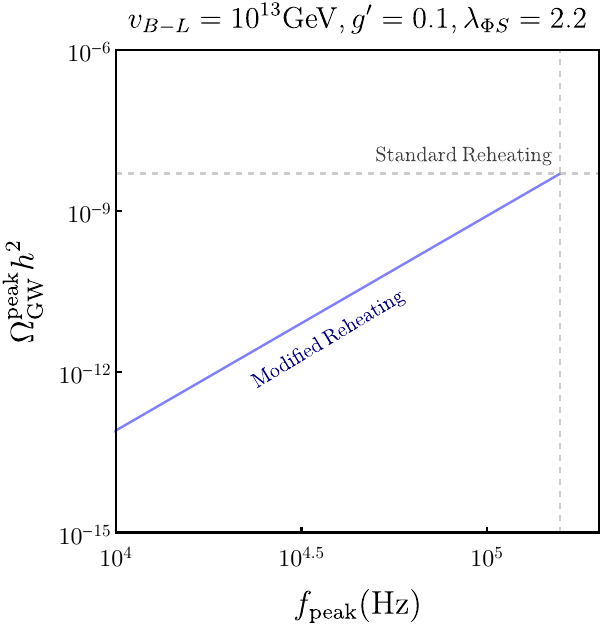}
 	\caption{\textit{Left:} The variation of $f_{\rm peak}$ with $M_N$ is indicated by the solid(dashed) line for modified(standard) reheating. \textit{Right:} The effect of varying $M_N$, similar to the left, but projected onto the $f_{\rm peak}$ vs $\Omega_{\rm GW}^{\rm peak}h^2$ plane.}\label{fig4}
 \end{figure}

For the singlet extension, however, $M_N$ can be lowered to much smaller values. As shown in Fig.\ \ref{fig3} left panel, this allows the RHN-mass–dependent matter-dominated era to persist even within the 3FL regime for $g^\prime=0.1$. The benchmark points BP2(${\spadesuit}$) and BP3(${\bigstar}$) correspond to nearly identical VEVs, but distinct values of $M_N$, representing the 3FL and 2FL regimes, respectively, while BP1(${\clubsuit}$) lies at both lower $M_N$ and lower VEV. The corresponding GW spectra in Fig.\ \ref{fig3} right panel clearly differentiate these regimes: in the 2FL case, entropy production is milder, whereas the 3FL case exhibits a longer matter-dominated phase due to the smaller Yukawa couplings, leading to enhanced spectral distortion (see Fig.\ \ref{fig4} right panel). As one moves toward lower $M_N$ and smaller VEV, the GW peak shifts more towards the lower frequencies (see Fig.\ \ref{fig4} left panel), rendering the 3FL regime phenomenologically more interesting for future observational tests.
The numerical results presented above highlight how the scalar reheating dynamics, governed by $M_N$ and $g^\prime$, shape both the early-Universe history and the resulting GW signal. Entropy injection from the scalar decay dilutes any preexisting baryon asymmetry when $M_N>T_{\rm dec}$, implying that viable scenarios favour $M_N<T_{\rm dec}$, where thermal leptogenesis proceeds efficiently after reheating. In cases with late decay, quasi-degenerate RHNs can circumvent entropy dilution through resonant leptogenesis, which justifies our choice of a single representative mass scale $M_N$.

Second, in recent years, there has been a growing effort to connect leptogenesis with primordial GWs, see, e.g., Refs.~\cite{Blasi:2020wpy,lepgw1,lepgw2,lepgw3,lepgw4,lepgw5,lepgw6,lepgw7,Ghoshal:2022kqp,lepgw8,Borah:2022cdx,Chun:2023ezg,Azatov:2021irb,Samanta:2025jec,Chianese:2024gee,Chianese:2024nyw,Chianese:2025mll,Liu:2025xvm,Ghosh:2025non}. This work complements those studies by showing that flavour regimes of leptogenesis can also be probed through GW signals generated during conformal first-order phase transitions. The connection arises because the RHN mass, which controls both the flavour dynamics and the reheating efficiency, leaves a characteristic imprint on the GW spectrum.

The amplitude and shape of the GW spectrum are strongly correlated with the efficiency of reheating: larger entropy production leads to more pronounced spectral suppression and a shift of the peak frequency to lower values. This mapping between microphysics and observables provides a distinctive test of conformal $U(1)_{B-L}$ dynamics. However, as discussed in Sec.\ \ref{s5}, the theoretical uncertainties in GW production—stemming from bubble dynamics, energy partitioning, and the modelling of sound-wave and turbulence sources—remain significant. Hence, the spectra shown here should be regarded as indicative templates rather than precise predictions. Future refinements incorporating improved calculations and numerical simulations will reduce these uncertainties, but the qualitative differences between flavour regimes are expected to remain.

Alternative mechanisms, such as bubble-driven leptogenesis with quasi-degenerate RHNs \cite{Dasgupta:2022isg}, may also operate if the lepton asymmetry is generated during the phase transition itself. In that case, the subsequent matter domination or reheating would still dilute the asymmetry, and since the transition scale exceeds the RHN mass, thermal leptogenesis would typically dominate the final baryon asymmetry.

In summary, our numerical analysis reveals that for $M_N<T_{\rm dec}$ the conformal $U(1)_{B-L}$ model can simultaneously accommodate successful thermal leptogenesis, nonthermal RHN dark matter, and distinctive GW spectral signatures. The dependence of GW spectra on the RHN mass and flavour regime highlights a new synergy between neutrino physics and GW cosmology, motivating future work incorporating a full baryon-asymmetry computation and a quantitative SNR-level detectability analysis.

\section{Conclusions \& Discussions}\label{s7}

We have studied the cosmological and phenomenological consequences of a supercooled first-order phase transition in the classically conformal $U(1)_{B-L}$ model, focusing on the interplay between leptogenesis, nonthermal RHN DM, and GW signatures. Our analysis demonstrates that the same scalar field responsible for symmetry breaking and RHN mass generation can also induce an EMD epoch through its slow decay into RHNs. The duration of this epoch, controlled by the RHN mass and the gauge coupling $g^\prime$, crucially determines the modification of the primordial GW spectrum by the entropy injection. 

Numerically, we find that for the minimal model with $g^\prime=0.5$, a scalar-dominated EMD epoch exists in a sizeable parameter region. This phase leaves a distinct, RHN-mass-dependent distortion in the GW spectrum, connecting the flavour structure of leptogenesis to potentially observable GW features. Smaller gauge couplings compress the viable parameter space and disfavour RHN-dominated EMD for $M_N\lesssim 10^9$ GeV, whereas singlet-extended realisations reopen this region, permitting even the 3FL leptogenesis regime to produce detectable GW signatures. 

The correlation between the reheating efficiency, the RHN mass, and the GW spectral shape provides a novel probe of the leptogenesis flavour regime that could be accessible to future GW observatories operating across the high-frequency regime. Future work will refine these results through a full computation of the baryon asymmetry and a detailed signal-to-noise analysis of the GW detectability.

\section{Acknowledgements}
The authors thank Yann Gouttenoire, Maciej Kierkla and Ke-Pan Xie for helpful discussions, and Rome Samanta for useful comments. The work of both PA, SD and ZZ is supported by the National Natural Science Foundation of China (NNSFC) under Key Projects grant No. 12335005.

\appendix
\section{Computation of the free-streaming length of RHN DM from scalar decay}\label{APP-A}
Assuming the $(B-L)$ scalar decays at rest in the comoving frame, the energy density of each RHN DM produced is
\begin{equation}
    E_{N_{\rm DM}}=\frac{m_\Phi}{2},
\end{equation}
since $m_\Phi\gg M_{\rm DM}$ during its decay, the initial momentum of $N_{\rm DM}$ is,
\begin{equation}
    p_{\rm DM}(t_{\rm dec})\sim  E_{N_{\rm DM}}.
\end{equation}
During cosmological evolution, the DM momentum redshifts as
\begin{equation}
    p_{\rm DM}(t)=p_{\rm DM}(t_{\rm dec}) \frac{a(t_{\rm dec})}{a(t)},
\end{equation}
which translates to the velocity
\begin{equation}
    v_{\rm DM}(t)=\frac{p_{\rm DM}(t)}{\sqrt{p_{\rm DM}^2(t)+M_{\rm DM}^2}}.
\end{equation}
The co-moving free-streaming length is
\begin{equation}
    \lambda_{\rm FS}=\int_{t_{\rm dec}}^{t_{\rm eq}}\frac{v_{\rm DM}(t)}{a(t)}dt,
\end{equation}
which can be written in terms of the scale factor
\begin{equation}
    \lambda_{\rm FS}=\int_{a_{\rm dec}}^{a_{\rm eq}} \frac{v_{\rm DM}(a)}{a^2 H(a)}da.
\end{equation}
Assuming radiation domination after the decay of $\Phi$ until matter radiation equality
\begin{equation}
    H(a)=H_0\sqrt{\Omega_R^0} a^{-2}.
\end{equation}
The DM velocity can be written in terms of the scale factor as
\begin{equation}
    v_{\rm DM}(a)=\frac{p_{\rm DM}(a)}{\sqrt{p_{\rm DM}^2(a)+M_{\rm DM}^2}}=\frac{\frac{m_\Phi}{2}\frac{a_{\rm dec}}{a}}{\sqrt{(\frac{m_\Phi}{2}\frac{a_{\rm dec}}{a})^2+M_{\rm DM}^2}}.
\end{equation}
Now letting $X=2M_{\rm DM}/(m_\Phi a_{\rm dec})$, the free streaming length can be simplified as,
\begin{equation}
    \lambda_{\rm FS}=\frac{1}{H_0\sqrt{\Omega_R^0}}\int_{a_{\rm dec}}^{a_{\rm eq}}\frac{da}{\sqrt{1+X^2 a^2}}.
    \end{equation}
    Since $a_{\rm dec}\ll a_{\rm eq}$, it can be approximated to
    \begin{equation}
        \lambda_{\rm FS}\sim \frac{1}{H_0\sqrt{\Omega_R^0} X}\sinh^{-1}(Xa_{\rm eq}).
    \end{equation}
    At matter-radiation equality, $\Omega_R^0a_{\rm eq}^{-4}=\Omega_M^0a_{\rm eq}^{-3}$, substituting this one gets
    \begin{equation}
        \lambda_{\rm FS}\sim \frac{\sqrt{\Omega_R^0}}{H_0\Omega_M^0} \frac{\sinh^{-1}(Z)}{Z},
    \end{equation}
    where $Z=a_{\rm eq}X$. Now using the Ly-$\alpha$ constraint $\lambda_{\rm FS}\lesssim1$ Mpc, with $H_0^{-1}=4.422\times 10^3$ Mpc, $\Omega_R^0=9\times 10^{-5}$, and $\Omega_M^0=0.3147$ from Planck 2018 data \cite{Planck:2018vyg}, a numerical estimate can be done,
    \begin{equation}
        Z\gtrsim 1015.25,
    \end{equation}
    which translates to
    \begin{equation}
        M_{\rm DM}\gtrsim 1015.25\left(\frac{g_{*s}^{\rm eq}}{g_{*s}^{\rm dec}} \right)^{1/3}\left( \frac{T_{\rm eq}}{T_{\rm dec}}\right)m_\Phi.
    \end{equation}
 \bibliography{main.bib}

@misc{Gonstal:2025qky,
    author = "Gonstal, Adam and Lewicki, Marek and Swiezewska, Bogumila",
    title = "{Reconstructing early universe evolution with gravitational waves from supercooled phase transitions}",
    eprint = "2502.18436",
    archivePrefix = "arXiv",
    primaryClass = "gr-qc",
    month = "2",
    year = "2025"
}

@article{Gouttenoire:2023pxh,
    author = "Gouttenoire, Yann",
    title = "{Primordial black holes from conformal Higgs}",
    eprint = "2311.13640",
    archivePrefix = "arXiv",
    primaryClass = "hep-ph",
    doi = "10.1016/j.physletb.2024.138800",
    journal = "Phys. Lett. B",
    volume = "855",
    pages = "138800",
    year = "2024"
}

@article{Gouttenoire:2023bqy,
    author = "Gouttenoire, Yann",
    title = "{First-Order Phase Transition Interpretation of Pulsar Timing Array Signal Is Consistent with Solar-Mass Black Holes}",
    eprint = "2307.04239",
    archivePrefix = "arXiv",
    primaryClass = "hep-ph",
    doi = "10.1103/PhysRevLett.131.171404",
    journal = "Phys. Rev. Lett.",
    volume = "131",
    number = "17",
    pages = "171404",
    year = "2023"
}

@article{Gouttenoire:2021kjv,
    author = "Gouttenoire, Yann and Jinno, Ryusuke and Sala, Filippo",
    title = "{Friction pressure on relativistic bubble walls}",
    eprint = "2112.07686",
    archivePrefix = "arXiv",
    primaryClass = "hep-ph",
    reportNumber = "DESY-21-147, IFT-UAM/CSIC-21-146",
    doi = "10.1007/JHEP05(2022)004",
    journal = "JHEP",
    volume = "05",
    pages = "004",
    year = "2022"
}

@article{Iso:2009ss,
    author = "Iso, Satoshi and Okada, Nobuchika and Orikasa, Yuta",
    title = "{Classically conformal $B^-$ L extended Standard Model}",
    eprint = "0902.4050",
    archivePrefix = "arXiv",
    primaryClass = "hep-ph",
    reportNumber = "KEK-TH-1303",
    doi = "10.1016/j.physletb.2009.04.046",
    journal = "Phys. Lett. B",
    volume = "676",
    pages = "81--87",
    year = "2009"
}

@article{Iso:2009nw,
    author = "Iso, Satoshi and Okada, Nobuchika and Orikasa, Yuta",
    title = "{The minimal B-L model naturally realized at TeV scale}",
    eprint = "0909.0128",
    archivePrefix = "arXiv",
    primaryClass = "hep-ph",
    reportNumber = "KEK-TH-1327",
    doi = "10.1103/PhysRevD.80.115007",
    journal = "Phys. Rev. D",
    volume = "80",
    pages = "115007",
    year = "2009"
}

@article{Das:2015nwk,
    author = "Das, Arindam and Okada, Nobuchika and Papapietro, Nathan",
    title = "{Electroweak vacuum stability in classically conformal B-L extension of the Standard Model}",
    eprint = "1509.01466",
    archivePrefix = "arXiv",
    primaryClass = "hep-ph",
    doi = "10.1140/epjc/s10052-017-4683-2",
    journal = "Eur. Phys. J. C",
    volume = "77",
    number = "2",
    pages = "122",
    year = "2017"
}

@article{Jinno:2016knw,
    author = "Jinno, Ryusuke and Takimoto, Masahiro",
    title = "{Probing a classically conformal B-L model with gravitational waves}",
    eprint = "1604.05035",
    archivePrefix = "arXiv",
    primaryClass = "hep-ph",
    reportNumber = "KEK-TH-1896",
    doi = "10.1103/PhysRevD.95.015020",
    journal = "Phys. Rev. D",
    volume = "95",
    number = "1",
    pages = "015020",
    year = "2017"
}

@article{Iso:2017uuu,
    author = "Iso, Satoshi and Serpico, Pasquale D. and Shimada, Kengo",
    title = "{QCD-Electroweak First-Order Phase Transition in a Supercooled Universe}",
    eprint = "1704.04955",
    archivePrefix = "arXiv",
    primaryClass = "hep-ph",
    reportNumber = "KEK-TH-1969, LAPTH-008-17",
    doi = "10.1103/PhysRevLett.119.141301",
    journal = "Phys. Rev. Lett.",
    volume = "119",
    number = "14",
    pages = "141301",
    year = "2017"
}

@article{Marzo:2018nov,
    author = "Marzo, Carlo and Marzola, Luca and Vaskonen, Ville",
    title = "{Phase transition and vacuum stability in the classically conformal B{\textendash}L model}",
    eprint = "1811.11169",
    archivePrefix = "arXiv",
    primaryClass = "hep-ph",
    reportNumber = "KCL-PH-TH/2018-68",
    doi = "10.1140/epjc/s10052-019-7076-x",
    journal = "Eur. Phys. J. C",
    volume = "79",
    number = "7",
    pages = "601",
    year = "2019"
}

@article{Ellis:2019oqb,
    author = "Ellis, John and Lewicki, Marek and No, Jos{\'e} Miguel and Vaskonen, Ville",
    title = "{Gravitational wave energy budget in strongly supercooled phase transitions}",
    eprint = "1903.09642",
    archivePrefix = "arXiv",
    primaryClass = "hep-ph",
    reportNumber = "KCL-PH-TH/2019-32, CERN-TH-2019-032, IFT-UAM/CSIC-19-32",
    doi = "10.1088/1475-7516/2019/06/024",
    journal = "JCAP",
    volume = "06",
    pages = "024",
    year = "2019"
}

@article{Ellis:2020nnr,
    author = "Ellis, John and Lewicki, Marek and Vaskonen, Ville",
    title = "{Updated predictions for gravitational waves produced in a strongly supercooled phase transition}",
    eprint = "2007.15586",
    archivePrefix = "arXiv",
    primaryClass = "astro-ph.CO",
    reportNumber = "KCL-PH-TH/2020-40, CERN-TH-2020-129",
    doi = "10.1088/1475-7516/2020/11/020",
    journal = "JCAP",
    volume = "11",
    pages = "020",
    year = "2020"
}

@article{Linde:1981zj,
    author = "Linde, Andrei D.",
    title = "{Decay of the False Vacuum at Finite Temperature}",
    reportNumber = "LEBEDEV-81-265",
    doi = "10.1016/0550-3213(83)90072-X",
    journal = "Nucl. Phys. B",
    volume = "216",
    pages = "421",
    year = "1983",
    note = "[Erratum: Nucl.Phys.B 223, 544 (1983)]"
}

@article{Ellis:2018mja,
    author = "Ellis, John and Lewicki, Marek and No, Jos{\'e} Miguel",
    title = "{On the Maximal Strength of a First-Order Electroweak Phase Transition and its Gravitational Wave Signal}",
    eprint = "1809.08242",
    archivePrefix = "arXiv",
    primaryClass = "hep-ph",
    reportNumber = "KCL-PH-TH/2018-46, CERN-TH/2018-197, IFT-UAM/CSIC-18-94, CERN-TH-2018-197",
    doi = "10.1088/1475-7516/2019/04/003",
    journal = "JCAP",
    volume = "04",
    pages = "003",
    year = "2019"
}

@article{Athron:2023xlk,
    author = "Athron, Peter and Bal{\'a}zs, Csaba and Fowlie, Andrew and Morris, Lachlan and Wu, Lei",
    title = "{Cosmological phase transitions: From perturbative particle physics to gravitational waves}",
    eprint = "2305.02357",
    archivePrefix = "arXiv",
    primaryClass = "hep-ph",
    doi = "10.1016/j.ppnp.2023.104094",
    journal = "Prog. Part. Nucl. Phys.",
    volume = "135",
    pages = "104094",
    year = "2024"
}

@article{Wainwright:2011kj,
    author = "Wainwright, Carroll L.",
    title = "{CosmoTransitions: Computing Cosmological Phase Transition Temperatures and Bubble Profiles with Multiple Fields}",
    eprint = "1109.4189",
    archivePrefix = "arXiv",
    primaryClass = "hep-ph",
    doi = "10.1016/j.cpc.2012.04.004",
    journal = "Comput. Phys. Commun.",
    volume = "183",
    pages = "2006--2013",
    year = "2012"
}

@article{Coleman:1977py,
    author = "Coleman, Sidney R.",
    title = "{The Fate of the False Vacuum. 1. Semiclassical Theory}",
    reportNumber = "HUTP-77-A004",
    doi = "10.1103/PhysRevD.16.1248",
    journal = "Phys. Rev. D",
    volume = "15",
    pages = "2929--2936",
    year = "1977",
    note = "[Erratum: Phys.Rev.D 16, 1248 (1977)]"
}

@article{Takahashi:2007tz,
    author = "Takahashi, Fuminobu",
    title = "{Gravitino dark matter from inflaton decay}",
    eprint = "0705.0579",
    archivePrefix = "arXiv",
    primaryClass = "hep-ph",
    reportNumber = "DESY-07-060",
    doi = "10.1016/j.physletb.2007.12.048",
    journal = "Phys. Lett. B",
    volume = "660",
    pages = "100--106",
    year = "2008"
}

@article{BhupalDev:2013oiy,
    author = "Bhupal Dev, P. S. and Mazumdar, Anupam and Qutub, Saleh",
    title = "{Constraining Non-thermal and Thermal properties of Dark Matter}",
    eprint = "1311.5297",
    archivePrefix = "arXiv",
    primaryClass = "hep-ph",
    reportNumber = "MAN-HEP-2013-23",
    doi = "10.3389/fphy.2014.00026",
    journal = "Front. in Phys.",
    volume = "2",
    pages = "26",
    year = "2014"
}

@article{Hisano:2000dz,
    author = "Hisano, Junji and Kohri, Kazunori and Nojiri, Mihoko M.",
    title = "{Neutralino warm dark matter}",
    eprint = "hep-ph/0011216",
    archivePrefix = "arXiv",
    reportNumber = "KEK-TH-725, CERN-TH-2000-338, YITP-00-62",
    doi = "10.1016/S0370-2693(01)00395-1",
    journal = "Phys. Lett. B",
    volume = "505",
    pages = "169--176",
    year = "2001"
}

@article{Planck:2018vyg,
    author = "Aghanim, N. and others",
    collaboration = "Planck",
    title = "{Planck 2018 results. VI. Cosmological parameters}",
    eprint = "1807.06209",
    archivePrefix = "arXiv",
    primaryClass = "astro-ph.CO",
    doi = "10.1051/0004-6361/201833910",
    journal = "Astron. Astrophys.",
    volume = "641",
    pages = "A6",
    year = "2020",
    note = "[Erratum: Astron.Astrophys. 652, C4 (2021)]"
}

@article{Enqvist:1991xw,
    author = "Enqvist, K. and Ignatius, J. and Kajantie, K. and Rummukainen, K.",
    title = "{Nucleation and bubble growth in a first order cosmological electroweak phase transition}",
    reportNumber = "HU-TFT-91-35",
    doi = "10.1103/PhysRevD.45.3415",
    journal = "Phys. Rev. D",
    volume = "45",
    pages = "3415--3428",
    year = "1992"
}

@article{Turner:1992tz,
    author = "Turner, Michael S. and Weinberg, Erick J. and Widrow, Lawrence M.",
    title = "{Bubble nucleation in first order inflation and other cosmological phase transitions}",
    reportNumber = "FERMILAB-PUB-91-334-A, CU-TP-558, IASSNS-HEP-92-21",
    doi = "10.1103/PhysRevD.46.2384",
    journal = "Phys. Rev. D",
    volume = "46",
    pages = "2384--2403",
    year = "1992"
}

@article{Athron:2022mmm,
    author = "Athron, Peter and Bal{\'a}zs, Csaba and Morris, Lachlan",
    title = "{Supercool subtleties of cosmological phase transitions}",
    eprint = "2212.07559",
    archivePrefix = "arXiv",
    primaryClass = "hep-ph",
    doi = "10.1088/1475-7516/2023/03/006",
    journal = "JCAP",
    volume = "03",
    pages = "006",
    year = "2023"
}

@article{seesaw4,
    author = "Mohapatra, R. N.",
    title = "{Mechanism for Understanding Small Neutrino Mass in Superstring Theories}",
    doi = "10.1103/PhysRevLett.56.561",
    journal = "Phys. Rev. Lett.",
    volume = "56",
    pages = "561--563",
    year = "1986"
}

@article{seesaw3,
    author = "Yanagida, Tsutomu",
    title = "{Horizontal Symmetry and Masses of Neutrinos}",
    reportNumber = "TU-80-208",
    doi = "10.1143/PTP.64.1103",
    journal = "Prog. Theor. Phys.",
    volume = "64",
    pages = "1103",
    year = "1980"
}

@article{seesaw2,
    author = "Gell-Mann, Murray and Ramond, Pierre and Slansky, Richard",
    title = "{Complex Spinors and Unified Theories}",
    eprint = "1306.4669",
    archivePrefix = "arXiv",
    primaryClass = "hep-th",
    reportNumber = "PRINT-80-0576",
    journal = "Conf. Proc. C",
    volume = "790927",
    pages = "315--321",
    year = "1979"
}

@article{seesaw1,
    author = "Minkowski, Peter",
    title = "{$\mu \to e\gamma$ at a Rate of One Out of $10^{9}$ Muon Decays?}",
    reportNumber = "Print-77-0182 (BERN)",
    doi = "10.1016/0370-2693(77)90435-X",
    journal = "Phys. Lett. B",
    volume = "67",
    pages = "421--428",
    year = "1977"
}

@article{fllep1,
    author = "Abada, A. and Davidson, S. and Ibarra, A. and Josse-Michaux, F. -X. and Losada, M. and Riotto, A.",
    title = "{Flavour Matters in Leptogenesis}",
    eprint = "hep-ph/0605281",
    archivePrefix = "arXiv",
    reportNumber = "CERN-PH-TH-2006-093, DNI-UAN-06-97FT, IFT-UAM-CSIC-06-23, LPT-ORSAY-06-21, LYCEN-2006-07",
    doi = "10.1088/1126-6708/2006/09/010",
    journal = "JHEP",
    volume = "09",
    pages = "010",
    year = "2006"
}

@article{fllep2,
    author = "Nardi, Enrico and Nir, Yosef and Roulet, Esteban and Racker, Juan",
    title = "{The Importance of flavor in leptogenesis}",
    eprint = "hep-ph/0601084",
    archivePrefix = "arXiv",
    doi = "10.1088/1126-6708/2006/01/164",
    journal = "JHEP",
    volume = "01",
    pages = "164",
    year = "2006"
}

@article{fllep3,
    author = "Blanchet, Steve and Di Bari, Pasquale",
    title = "{Flavor effects on leptogenesis predictions}",
    eprint = "hep-ph/0607330",
    archivePrefix = "arXiv",
    doi = "10.1088/1475-7516/2007/03/018",
    journal = "JCAP",
    volume = "03",
    pages = "018",
    year = "2007"
}

@article{Blasi:2020wpy,
    author = "Blasi, Simone and Brdar, Vedran and Schmitz, Kai",
    title = "{Fingerprint of low-scale leptogenesis in the primordial gravitational-wave spectrum}",
    eprint = "2004.02889",
    archivePrefix = "arXiv",
    primaryClass = "hep-ph",
    reportNumber = "CERN-TH-2020-055",
    doi = "10.1103/PhysRevResearch.2.043321",
    journal = "Phys. Rev. Res.",
    volume = "2",
    number = "4",
    pages = "043321",
    year = "2020"
}

@article{lepgw7,
    author = "Perez-Gonzalez, Yuber F. and Turner, Jessica",
    title = "{Assessing the tension between a black hole dominated early universe and leptogenesis}",
    eprint = "2010.03565",
    archivePrefix = "arXiv",
    primaryClass = "hep-ph",
    reportNumber = "FERMILAB-PUB-20-528-T, NUHEP-TH/20-10, IPPP/20/46",
    doi = "10.1103/PhysRevD.104.103021",
    journal = "Phys. Rev. D",
    volume = "104",
    number = "10",
    pages = "103021",
    year = "2021"
}

@article{lepgw6,
    author = "Borah, Debasish and Jyoti Das, Suruj and Samanta, Rome and Urban, Federico R.",
    title = "{PBH-infused seesaw origin of matter and unique gravitational waves}",
    eprint = "2211.15726",
    archivePrefix = "arXiv",
    primaryClass = "hep-ph",
    doi = "10.1007/JHEP03(2023)127",
    journal = "JHEP",
    volume = "03",
    pages = "127",
    year = "2023"
}

@article{lepgw5,
    author = "Dasgupta, Arnab and Dev, P. S. Bhupal and Ghoshal, Anish and Mazumdar, Anupam",
    title = "{Gravitational wave pathway to testable leptogenesis}",
    eprint = "2206.07032",
    archivePrefix = "arXiv",
    primaryClass = "hep-ph",
    doi = "10.1103/PhysRevD.106.075027",
    journal = "Phys. Rev. D",
    volume = "106",
    number = "7",
    pages = "075027",
    year = "2022"
}

@article{lepgw4,
    author = "Barman, Basabendu and Borah, Debasish and Dasgupta, Arnab and Ghoshal, Anish",
    title = "{Probing high scale Dirac leptogenesis via gravitational waves from domain walls}",
    eprint = "2205.03422",
    archivePrefix = "arXiv",
    primaryClass = "hep-ph",
    doi = "10.1103/PhysRevD.106.015007",
    journal = "Phys. Rev. D",
    volume = "106",
    number = "1",
    pages = "015007",
    year = "2022"
}

@article{lepgw3,
    author = "Datta, Satyabrata and Ghosal, Ambar and Samanta, Rome",
    title = "{Baryogenesis from ultralight primordial black holes and strong gravitational waves from cosmic strings}",
    eprint = "2012.14981",
    archivePrefix = "arXiv",
    primaryClass = "hep-ph",
    doi = "10.1088/1475-7516/2021/08/021",
    journal = "JCAP",
    volume = "08",
    pages = "021",
    year = "2021"
}

@article{lepgw2,
    author = "Samanta, Rome and Datta, Satyabrata",
    title = "{Gravitational wave complementarity and impact of NANOGrav data on gravitational leptogenesis}",
    eprint = "2009.13452",
    archivePrefix = "arXiv",
    primaryClass = "hep-ph",
    doi = "10.1007/JHEP05(2021)211",
    journal = "JHEP",
    volume = "05",
    pages = "211",
    year = "2021"
}

@article{lepgw1,
    author = "Dror, Jeff A. and Hiramatsu, Takashi and Kohri, Kazunori and Murayama, Hitoshi and White, Graham",
    title = "{Testing the Seesaw Mechanism and Leptogenesis with Gravitational Waves}",
    eprint = "1908.03227",
    archivePrefix = "arXiv",
    primaryClass = "hep-ph",
    reportNumber = "IPMU19-0108, DESY-19-138, DESY 19-138, KEK-TH-2147, KEK-Cosmo-241",
    doi = "10.1103/PhysRevLett.124.041804",
    journal = "Phys. Rev. Lett.",
    volume = "124",
    number = "4",
    pages = "041804",
    year = "2020"
}

@article{Borah:2022cdx,
    author = "Borah, Debasish and Dasgupta, Arnab and Saha, Indrajit",
    title = "{Leptogenesis and dark matter through relativistic bubble walls with observable gravitational waves}",
    eprint = "2207.14226",
    archivePrefix = "arXiv",
    primaryClass = "hep-ph",
    doi = "10.1007/JHEP11(2022)136",
    journal = "JHEP",
    volume = "11",
    pages = "136",
    year = "2022"
}

@article{lepgw8,
    author = "Huang, Peisi and Xie, Ke-Pan",
    title = "{Leptogenesis triggered by a first-order phase transition}",
    eprint = "2206.04691",
    archivePrefix = "arXiv",
    primaryClass = "hep-ph",
    doi = "10.1007/JHEP09(2022)052",
    journal = "JHEP",
    volume = "09",
    pages = "052",
    year = "2022"
}

@article{Ghoshal:2022kqp,
    author = "Ghoshal, Anish and Samanta, Rome and White, Graham",
    title = "{Bremsstrahlung high-frequency gravitational wave signatures of high-scale nonthermal leptogenesis}",
    eprint = "2211.10433",
    archivePrefix = "arXiv",
    primaryClass = "hep-ph",
    doi = "10.1103/PhysRevD.108.035019",
    journal = "Phys. Rev. D",
    volume = "108",
    number = "3",
    pages = "035019",
    year = "2023"
}

@article{Chun:2023ezg,
    author = "Chun, Eung Jin and Dutka, Tomasz P. and Jung, Tae Hyun and Nagels, Xander and Vanvlasselaer, Miguel",
    title = "{Bubble-assisted leptogenesis}",
    eprint = "2305.10759",
    archivePrefix = "arXiv",
    primaryClass = "hep-ph",
    reportNumber = "CTPU-PTC-23-17",
    doi = "10.1007/JHEP09(2023)164",
    journal = "JHEP",
    volume = "09",
    pages = "164",
    year = "2023"
}

@article{Azatov:2021irb,
    author = "Azatov, Aleksandr and Vanvlasselaer, Miguel and Yin, Wen",
    title = "{Baryogenesis via relativistic bubble walls}",
    eprint = "2106.14913",
    archivePrefix = "arXiv",
    primaryClass = "hep-ph",
    reportNumber = "SISSA 13/2021/FISI TU-1127",
    doi = "10.1007/JHEP10(2021)043",
    journal = "JHEP",
    volume = "10",
    pages = "043",
    year = "2021"
}

@misc{Chianese:2025mll,
    author = "Chianese, Marco and Dom{\`e}nech, Guillem and Papanikolaou, Theodoros and Samanta, Rome and Saviano, Ninetta",
    title = "{Induced Gravitational Waves as Cosmic Tracers of Leptogenesis}",
    eprint = "2504.20135",
    archivePrefix = "arXiv",
    primaryClass = "hep-ph",
    month = "4",
    year = "2025"
}

@article{Samanta:2025jec,
    author = "Samanta, Rome",
    title = "{Probing leptogenesis at LISA: a Fisher analysis}",
    eprint = "2503.09884",
    archivePrefix = "arXiv",
    primaryClass = "hep-ph",
    doi = "10.1088/1475-7516/2025/08/095",
    journal = "JCAP",
    volume = "08",
    pages = "095",
    year = "2025"
}

@article{Chianese:2024gee,
    author = "Chianese, Marco and Datta, Satyabrata and Miele, Gennaro and Samanta, Rome and Saviano, Ninetta",
    title = "{Probing flavored regimes of leptogenesis with gravitational waves from cosmic strings}",
    eprint = "2406.01231",
    archivePrefix = "arXiv",
    primaryClass = "hep-ph",
    doi = "10.1103/PhysRevD.111.L041305",
    journal = "Phys. Rev. D",
    volume = "111",
    number = "4",
    pages = "L041305",
    year = "2025"
}

@article{Chianese:2024nyw,
    author = "Chianese, Marco and Datta, Satyabrata and Samanta, Rome and Saviano, Ninetta",
    title = "{Tomography of flavoured leptogenesis with primordial blue gravitational waves}",
    eprint = "2405.00641",
    archivePrefix = "arXiv",
    primaryClass = "hep-ph",
    doi = "10.1088/1475-7516/2024/11/051",
    journal = "JCAP",
    volume = "11",
    pages = "051",
    year = "2024"
}

@article{Konstandin:2017sat,
    author = "Konstandin, Thomas",
    title = "{Gravitational radiation from a bulk flow model}",
    eprint = "1712.06869",
    archivePrefix = "arXiv",
    primaryClass = "astro-ph.CO",
    reportNumber = "DESY-17-227",
    doi = "10.1088/1475-7516/2018/03/047",
    journal = "JCAP",
    volume = "03",
    pages = "047",
    year = "2018"
}

@article{Lewicki:2020azd,
    author = "Lewicki, Marek and Vaskonen, Ville",
    title = "{Gravitational waves from colliding vacuum bubbles in gauge theories}",
    eprint = "2012.07826",
    archivePrefix = "arXiv",
    primaryClass = "astro-ph.CO",
    doi = "10.1140/epjc/s10052-021-09232-3",
    journal = "Eur. Phys. J. C",
    volume = "81",
    number = "5",
    pages = "437",
    year = "2021",
    note = "[Erratum: Eur.Phys.J.C 81, 1077 (2021)]"
}

@article{Cutting:2020nla,
    author = "Cutting, Daniel and Escartin, Elba Granados and Hindmarsh, Mark and Weir, David J.",
    title = "{Gravitational waves from vacuum first order phase transitions II: from thin to thick walls}",
    eprint = "2005.13537",
    archivePrefix = "arXiv",
    primaryClass = "astro-ph.CO",
    reportNumber = "HIP-2020-13/TH",
    doi = "10.1103/PhysRevD.103.023531",
    journal = "Phys. Rev. D",
    volume = "103",
    number = "2",
    pages = "023531",
    year = "2021"
}

@article{Durrer:2003ja,
    author = "Durrer, Ruth and Caprini, Chiara",
    title = "{Primordial magnetic fields and causality}",
    eprint = "astro-ph/0305059",
    archivePrefix = "arXiv",
    doi = "10.1088/1475-7516/2003/11/010",
    journal = "JCAP",
    volume = "11",
    pages = "010",
    year = "2003"
}

@article{Caprini:2009fx,
    author = "Caprini, Chiara and Durrer, Ruth and Konstandin, Thomas and Servant, Geraldine",
    title = "{General Properties of the Gravitational Wave Spectrum from Phase Transitions}",
    eprint = "0901.1661",
    archivePrefix = "arXiv",
    primaryClass = "astro-ph.CO",
    doi = "10.1103/PhysRevD.79.083519",
    journal = "Phys. Rev. D",
    volume = "79",
    pages = "083519",
    year = "2009"
}

@article{Cai:2019cdl,
    author = "Cai, Rong-Gen and Pi, Shi and Sasaki, Misao",
    title = "{Universal infrared scaling of gravitational wave background spectra}",
    eprint = "1909.13728",
    archivePrefix = "arXiv",
    primaryClass = "astro-ph.CO",
    reportNumber = "IPMU19-0135, YITP-19-88",
    doi = "10.1103/PhysRevD.102.083528",
    journal = "Phys. Rev. D",
    volume = "102",
    number = "8",
    pages = "083528",
    year = "2020"
}

@article{Barenboim:2016mjm,
    author = "Barenboim, Gabriela and Park, Wan-Il",
    title = "{Gravitational waves from first order phase transitions as a probe of an early matter domination era and its inverse problem}",
    eprint = "1605.03781",
    archivePrefix = "arXiv",
    primaryClass = "astro-ph.CO",
    reportNumber = "FTUV-16-05-07, IFIC-16-25",
    doi = "10.1016/j.physletb.2016.06.009",
    journal = "Phys. Lett. B",
    volume = "759",
    pages = "430--438",
    year = "2016"
}

@article{Dasgupta:2022isg,
    author = "Dasgupta, Arnab and Dev, P. S. Bhupal and Ghoshal, Anish and Mazumdar, Anupam",
    title = "{Gravitational wave pathway to testable leptogenesis}",
    eprint = "2206.07032",
    archivePrefix = "arXiv",
    primaryClass = "hep-ph",
    doi = "10.1103/PhysRevD.106.075027",
    journal = "Phys. Rev. D",
    volume = "106",
    number = "7",
    pages = "075027",
    year = "2022"
}

@article{Huang:2022vkf,
    author = "Huang, Peisi and Xie, Ke-Pan",
    title = "{Leptogenesis triggered by a first-order phase transition}",
    eprint = "2206.04691",
    archivePrefix = "arXiv",
    primaryClass = "hep-ph",
    doi = "10.1007/JHEP09(2022)052",
    journal = "JHEP",
    volume = "09",
    pages = "052",
    year = "2022"
}

@article{Bian:2019szo,
    author = "Bian, Ligong and Cheng, Wei and Guo, Huai-Ke and Zhang, Yongchao",
    title = "{Cosmological implications of a B {\ensuremath{-}} L charged hidden scalar: leptogenesis and gravitational waves}",
    eprint = "1907.13589",
    archivePrefix = "arXiv",
    primaryClass = "hep-ph",
    doi = "10.1088/1674-1137/ac1e09",
    journal = "Chin. Phys. C",
    volume = "45",
    number = "11",
    pages = "113104",
    year = "2021"
}

@article{Bodeker:2009qy,
    author = "Bodeker, Dietrich and Moore, Guy D.",
    title = "{Can electroweak bubble walls run away?}",
    eprint = "0903.4099",
    archivePrefix = "arXiv",
    primaryClass = "hep-ph",
    doi = "10.1088/1475-7516/2009/05/009",
    journal = "JCAP",
    volume = "05",
    pages = "009",
    year = "2009"
}

@article{Lewicki:2022pdb,
    author = "Lewicki, Marek and Vaskonen, Ville",
    title = "{Gravitational waves from bubble collisions and fluid motion in strongly supercooled phase transitions}",
    eprint = "2208.11697",
    archivePrefix = "arXiv",
    primaryClass = "astro-ph.CO",
    doi = "10.1140/epjc/s10052-023-11241-3",
    journal = "Eur. Phys. J. C",
    volume = "83",
    number = "2",
    pages = "109",
    year = "2023"
}

@article{Athron:2020sbe,
    author = "Athron, Peter and Bal{\'a}zs, Csaba and Fowlie, Andrew and Zhang, Yang",
    title = "{PhaseTracer: tracing cosmological phases and calculating transition properties}",
    eprint = "2003.02859",
    archivePrefix = "arXiv",
    primaryClass = "hep-ph",
    reportNumber = "CoEPP-MN-20-3",
    doi = "10.1140/epjc/s10052-020-8035-2",
    journal = "Eur. Phys. J. C",
    volume = "80",
    number = "6",
    pages = "567",
    year = "2020"
}

@article{Athron:2024xrh,
    author = "Athron, Peter and Balazs, Csaba and Fowlie, Andrew and Morris, Lachlan and Searle, William and Xiao, Yang and Zhang, Yang",
    title = "{PhaseTracer2: from the effective potential to gravitational waves}",
    eprint = "2412.04881",
    archivePrefix = "arXiv",
    primaryClass = "astro-ph.CO",
    doi = "10.1140/epjc/s10052-025-14258-y",
    journal = "Eur. Phys. J. C",
    volume = "85",
    number = "5",
    pages = "559",
    year = "2025"
}

@article{Fukugita:1986hr,
    author = "Fukugita, M. and Yanagida, T.",
    title = "{Baryogenesis Without Grand Unification}",
    reportNumber = "RIFP-641",
    doi = "10.1016/0370-2693(86)91126-3",
    journal = "Phys. Lett. B",
    volume = "174",
    pages = "45--47",
    year = "1986"
}

@article{Pilaftsis:2003gt,
    author = "Pilaftsis, Apostolos and Underwood, Thomas E. J.",
    title = "{Resonant leptogenesis}",
    eprint = "hep-ph/0309342",
    archivePrefix = "arXiv",
    reportNumber = "MC-TH-2003-09",
    doi = "10.1016/j.nuclphysb.2004.05.029",
    journal = "Nucl. Phys. B",
    volume = "692",
    pages = "303--345",
    year = "2004"
}

@article{Buchmuller:2004nz,
    author = "Buchmuller, W. and Di Bari, P. and Plumacher, M.",
    title = "{Leptogenesis for pedestrians}",
    eprint = "hep-ph/0401240",
    archivePrefix = "arXiv",
    reportNumber = "DESY-03-100, UAB-FT-551, CERN-TH-2003-199",
    doi = "10.1016/j.aop.2004.02.003",
    journal = "Annals Phys.",
    volume = "315",
    pages = "305--351",
    year = "2005"
}

@article{Davidson:2008bu,
    author = "Davidson, Sacha and Nardi, Enrico and Nir, Yosef",
    title = "{Leptogenesis}",
    eprint = "0802.2962",
    archivePrefix = "arXiv",
    primaryClass = "hep-ph",
    doi = "10.1016/j.physrep.2008.06.002",
    journal = "Phys. Rept.",
    volume = "466",
    pages = "105--177",
    year = "2008"
}

@article{Riotto:1999yt,
    author = "Riotto, Antonio and Trodden, Mark",
    title = "{Recent progress in baryogenesis}",
    eprint = "hep-ph/9901362",
    archivePrefix = "arXiv",
    reportNumber = "CERN-TH-99-04, CWRU-P6-99",
    doi = "10.1146/annurev.nucl.49.1.35",
    journal = "Ann. Rev. Nucl. Part. Sci.",
    volume = "49",
    pages = "35--75",
    year = "1999"
}

@article{uni1,
    author = "Asaka, Takehiko and Blanchet, Steve and Shaposhnikov, Mikhail",
    title = "{The nuMSM, dark matter and neutrino masses}",
    eprint = "hep-ph/0503065",
    archivePrefix = "arXiv",
    doi = "10.1016/j.physletb.2005.09.070",
    journal = "Phys. Lett. B",
    volume = "631",
    pages = "151--156",
    year = "2005"
}

@article{uni2,
    author = "Asaka, Takehiko and Blanchet, Steve and Shaposhnikov, Mikhail",
    title = "{The nuMSM, dark matter and neutrino masses}",
    eprint = "hep-ph/0503065",
    archivePrefix = "arXiv",
    doi = "10.1016/j.physletb.2005.09.070",
    journal = "Phys. Lett. B",
    volume = "631",
    pages = "151--156",
    year = "2005"
}

@article{uni3,
    author = "Dodelson, Scott and Widrow, Lawrence M.",
    title = "{Sterile-neutrinos as dark matter}",
    eprint = "hep-ph/9303287",
    archivePrefix = "arXiv",
    reportNumber = "FERMILAB-PUB-93-057-A",
    doi = "10.1103/PhysRevLett.72.17",
    journal = "Phys. Rev. Lett.",
    volume = "72",
    pages = "17--20",
    year = "1994"
}

@article{uni4,
    author = "Shi, Xiang-Dong and Fuller, George M.",
    title = "{A New dark matter candidate: Nonthermal sterile neutrinos}",
    eprint = "astro-ph/9810076",
    archivePrefix = "arXiv",
    doi = "10.1103/PhysRevLett.82.2832",
    journal = "Phys. Rev. Lett.",
    volume = "82",
    pages = "2832--2835",
    year = "1999"
}

@article{uni5,
    author = "Di Bari, Pasquale and Ludl, Patrick Otto and Palomares-Ruiz, Sergio",
    title = "{Unifying leptogenesis, dark matter and high-energy neutrinos with right-handed neutrino mixing via Higgs portal}",
    eprint = "1606.06238",
    archivePrefix = "arXiv",
    primaryClass = "hep-ph",
    reportNumber = "IFIC-16-42, IFIC/16-42",
    doi = "10.1088/1475-7516/2016/11/044",
    journal = "JCAP",
    volume = "11",
    pages = "044",
    year = "2016"
}

@article{uni6,
    author = "Di Bari, P. and Farrag, K. and Samanta, R. and Zhou, Y. L.",
    title = "{Density matrix calculation of the dark matter abundance in the Higgs induced right-handed neutrino mixing model}",
    eprint = "1908.00521",
    archivePrefix = "arXiv",
    primaryClass = "hep-ph",
    doi = "10.1088/1475-7516/2020/10/029",
    journal = "JCAP",
    volume = "10",
    pages = "029",
    year = "2020"
}

@article{uni7,
    author = "Datta, Arghyajit and Roshan, Rishav and Sil, Arunansu",
    title = "{Imprint of the Seesaw Mechanism on Feebly Interacting Dark Matter and the Baryon Asymmetry}",
    eprint = "2104.02030",
    archivePrefix = "arXiv",
    primaryClass = "hep-ph",
    doi = "10.1103/PhysRevLett.127.231801",
    journal = "Phys. Rev. Lett.",
    volume = "127",
    number = "23",
    pages = "231801",
    year = "2021"
}

@article{Athron:2023rfq,
    author = "Athron, Peter and Morris, Lachlan and Xu, Zhongxiu",
    title = "{How robust are gravitational wave predictions from cosmological phase transitions?}",
    eprint = "2309.05474",
    archivePrefix = "arXiv",
    primaryClass = "hep-ph",
    doi = "10.1088/1475-7516/2024/05/075",
    journal = "JCAP",
    volume = "05",
    pages = "075",
    year = "2024"
}

@article{Athron:2022jyi,
    author = "Athron, Peter and Balazs, Csaba and Fowlie, Andrew and Morris, Lachlan and White, Graham and Zhang, Yang",
    title = "{How arbitrary are perturbative calculations of the electroweak phase transition?}",
    eprint = "2208.01319",
    archivePrefix = "arXiv",
    primaryClass = "hep-ph",
    doi = "10.1007/JHEP01(2023)050",
    journal = "JHEP",
    volume = "01",
    pages = "050",
    year = "2023"
}

@article{Croon:2020cgk,
    author = "Croon, Djuna and Gould, Oliver and Schicho, Philipp and Tenkanen, Tuomas V. I. and White, Graham",
    title = "{Theoretical uncertainties for cosmological first-order phase transitions}",
    eprint = "2009.10080",
    archivePrefix = "arXiv",
    primaryClass = "hep-ph",
    reportNumber = "HIP-2020-26/TH",
    doi = "10.1007/JHEP04(2021)055",
    journal = "JHEP",
    volume = "04",
    pages = "055",
    year = "2021"
}

@article{Jinno:2017fby,
    author = "Jinno, Ryusuke and Takimoto, Masahiro",
    title = "{Gravitational waves from bubble dynamics: Beyond the Envelope}",
    eprint = "1707.03111",
    archivePrefix = "arXiv",
    primaryClass = "hep-ph",
    reportNumber = "CTPU-17-26, KEK-TH-1986",
    doi = "10.1088/1475-7516/2019/01/060",
    journal = "JCAP",
    volume = "01",
    pages = "060",
    year = "2019"
}

@article{Hindmarsh:2017gnf,
    author = "Hindmarsh, Mark and Huber, Stephan J. and Rummukainen, Kari and Weir, David J.",
    title = "{Shape of the acoustic gravitational wave power spectrum from a first order phase transition}",
    eprint = "1704.05871",
    archivePrefix = "arXiv",
    primaryClass = "astro-ph.CO",
    reportNumber = "HIP-2017-02-TH, HIP-2017-02/TH",
    doi = "10.1103/PhysRevD.96.103520",
    journal = "Phys. Rev. D",
    volume = "96",
    number = "10",
    pages = "103520",
    year = "2017",
    note = "[Erratum: Phys.Rev.D 101, 089902 (2020)]"
}

@article{Hindmarsh:2016lnk,
    author = "Hindmarsh, Mark",
    title = "{Sound shell model for acoustic gravitational wave production at a first-order phase transition in the early Universe}",
    eprint = "1608.04735",
    archivePrefix = "arXiv",
    primaryClass = "astro-ph.CO",
    doi = "10.1103/PhysRevLett.120.071301",
    journal = "Phys. Rev. Lett.",
    volume = "120",
    number = "7",
    pages = "071301",
    year = "2018"
}

@article{Hindmarsh:2019phv,
    author = "Hindmarsh, Mark and Hijazi, Mulham",
    title = "{Gravitational waves from first order cosmological phase transitions in the Sound Shell Model}",
    eprint = "1909.10040",
    archivePrefix = "arXiv",
    primaryClass = "astro-ph.CO",
    reportNumber = "NORDITA-2019-083, HIP-2019-29/TH",
    doi = "10.1088/1475-7516/2019/12/062",
    journal = "JCAP",
    volume = "12",
    pages = "062",
    year = "2019"
}

@article{ParticleDataGroup:2020ssz,
    author = "Zyla, P. A. and others",
    collaboration = "Particle Data Group",
    title = "{Review of Particle Physics}",
    doi = "10.1093/ptep/ptaa104",
    journal = "PTEP",
    volume = "2020",
    number = "8",
    pages = "083C01",
    year = "2020"
}

@article{Weinberg:1979bt,
    author = "Weinberg, Steven",
    title = "{Cosmological Production of Baryons}",
    reportNumber = "HUTP-78/A040",
    doi = "10.1103/PhysRevLett.42.850",
    journal = "Phys. Rev. Lett.",
    volume = "42",
    pages = "850--853",
    year = "1979"
}

@article{Kolb:1979qa,
    author = "Kolb, Edward W. and Wolfram, Stephen",
    title = "{Baryon Number Generation in the Early Universe}",
    reportNumber = "Print-79-0956 (CAL TECH), OAP-579, CALT-68-754",
    doi = "10.1016/0550-3213(82)90012-8",
    journal = "Nucl. Phys. B",
    volume = "172",
    pages = "224",
    year = "1980",
    note = "[Erratum: Nucl.Phys.B 195, 542 (1982)]"
}

@article{Okada:2012sg,
    author = "Okada, Nobuchika and Orikasa, Yuta",
    title = "{Dark matter in the classically conformal B-L model}",
    eprint = "1202.1405",
    archivePrefix = "arXiv",
    primaryClass = "hep-ph",
    reportNumber = "KEK-TH-1518",
    doi = "10.1103/PhysRevD.85.115006",
    journal = "Phys. Rev. D",
    volume = "85",
    pages = "115006",
    year = "2012"
}

@article{Khoze:2013oga,
    author = "Khoze, Valentin V. and Ro, Gunnar",
    title = "{Leptogenesis and Neutrino Oscillations in the Classically Conformal Standard Model with the Higgs Portal}",
    eprint = "1307.3764",
    archivePrefix = "arXiv",
    primaryClass = "hep-ph",
    reportNumber = "DCPT-13-102, IPPP-13-51",
    doi = "10.1007/JHEP10(2013)075",
    journal = "JHEP",
    volume = "10",
    pages = "075",
    year = "2013"
}

@article{Oda:2015gna,
    author = "Oda, Satsuki and Okada, Nobuchika and Takahashi, Dai-suke",
    title = "{Classically conformal U(1)' extended standard model and Higgs vacuum stability}",
    eprint = "1504.06291",
    archivePrefix = "arXiv",
    primaryClass = "hep-ph",
    doi = "10.1103/PhysRevD.92.015026",
    journal = "Phys. Rev. D",
    volume = "92",
    number = "1",
    pages = "015026",
    year = "2015"
}

@article{Das:2016zue,
    author = "Das, Arindam and Oda, Satsuki and Okada, Nobuchika and Takahashi, Dai-suke",
    title = "{Classically conformal U(1)' extended standard model, electroweak vacuum stability, and LHC Run-2 bounds}",
    eprint = "1605.01157",
    archivePrefix = "arXiv",
    primaryClass = "hep-ph",
    doi = "10.1103/PhysRevD.93.115038",
    journal = "Phys. Rev. D",
    volume = "93",
    number = "11",
    pages = "115038",
    year = "2016"
}

@article{Mohapatra:2023aei,
    author = "Mohapatra, Rabindra N. and Okada, Nobuchika",
    title = "{Conformal B-L and pseudo-Goldstone dark matter}",
    eprint = "2302.11072",
    archivePrefix = "arXiv",
    primaryClass = "hep-ph",
    doi = "10.1103/PhysRevD.107.095023",
    journal = "Phys. Rev. D",
    volume = "107",
    number = "9",
    pages = "095023",
    year = "2023"
}

@article{Coleman:1973jx,
    author = "Coleman, Sidney R. and Weinberg, Erick J.",
    title = "{Radiative Corrections as the Origin of Spontaneous Symmetry Breaking}",
    doi = "10.1103/PhysRevD.7.1888",
    journal = "Phys. Rev. D",
    volume = "7",
    pages = "1888--1910",
    year = "1973"
}

@article{Humbert:2015epa,
    author = "Humbert, Pascal and Lindner, Manfred and Smirnov, Juri",
    title = "{The Inverse Seesaw in Conformal Electro-Weak Symmetry Breaking and Phenomenological Consequences}",
    eprint = "1503.03066",
    archivePrefix = "arXiv",
    primaryClass = "hep-ph",
    doi = "10.1007/JHEP06(2015)035",
    journal = "JHEP",
    volume = "06",
    pages = "035",
    year = "2015"
}

@article{Lindner:2014oea,
    author = "Lindner, Manfred and Schmidt, Steffen and Smirnov, Juri",
    title = "{Neutrino Masses and Conformal Electro-Weak Symmetry Breaking}",
    eprint = "1405.6204",
    archivePrefix = "arXiv",
    primaryClass = "hep-ph",
    doi = "10.1007/JHEP10(2014)177",
    journal = "JHEP",
    volume = "10",
    pages = "177",
    year = "2014"
}

@article{Khoze:2013uia,
    author = "Khoze, Valentin V.",
    title = "{Inflation and Dark Matter in the Higgs Portal of Classically Scale Invariant Standard Model}",
    eprint = "1308.6338",
    archivePrefix = "arXiv",
    primaryClass = "hep-ph",
    reportNumber = "IPPP-13-65, DCPT-13-130",
    doi = "10.1007/JHEP11(2013)215",
    journal = "JHEP",
    volume = "11",
    pages = "215",
    year = "2013"
}

@article{Guth:1980zk,
    author = "Guth, Alan H. and Weinberg, Erick J.",
    title = "{A Cosmological Lower Bound on the Higgs Boson Mass}",
    reportNumber = "SLAC-PUB-2525",
    doi = "10.1103/PhysRevLett.45.1131",
    journal = "Phys. Rev. Lett.",
    volume = "45",
    pages = "1131",
    year = "1980"
}

@article{Witten:1980ez,
    author = "Witten, Edward",
    title = "{Cosmological Consequences of a Light Higgs Boson}",
    reportNumber = "HUTP-80/A040",
    doi = "10.1016/0550-3213(81)90182-6",
    journal = "Nucl. Phys. B",
    volume = "177",
    pages = "477--488",
    year = "1981"
}

@article{Hambye:2013dgv,
    author = "Hambye, Thomas and Strumia, Alessandro",
    title = "{Dynamical generation of the weak and Dark Matter scale}",
    eprint = "1306.2329",
    archivePrefix = "arXiv",
    primaryClass = "hep-ph",
    doi = "10.1103/PhysRevD.88.055022",
    journal = "Phys. Rev. D",
    volume = "88",
    pages = "055022",
    year = "2013"
}

@article{Lewicki:2024sfw,
    author = "Lewicki, Marek and Toczek, Piotr and Vaskonen, Ville",
    title = "{Black holes and gravitational waves from phase transitions in realistic models}",
    eprint = "2412.10366",
    archivePrefix = "arXiv",
    primaryClass = "astro-ph.CO",
    doi = "10.1016/j.dark.2025.102075",
    journal = "Phys. Dark Univ.",
    volume = "50",
    pages = "102075",
    year = "2025"
}

@article{Goncalves:2025uwh,
    author = "Gon{\c{c}}alves, Jo{\~a}o and Marfatia, Danny and Morais, Ant{\'o}nio P. and Pasechnik, Roman",
    title = "{Supercooled phase transitions in conformal dark sectors explain NANOGrav data}",
    eprint = "2501.11619",
    archivePrefix = "arXiv",
    primaryClass = "hep-ph",
    doi = "10.1016/j.physletb.2025.139829",
    journal = "Phys. Lett. B",
    volume = "869",
    pages = "139829",
    year = "2025"
}

@article{Li:2025nja,
    author = "Li, Jinzheng and Nath, Pran",
    title = "{Supercooled phase transitions: Why thermal history of hidden sector matters in analysis of pulsar timing array signals}",
    eprint = "2501.14986",
    archivePrefix = "arXiv",
    primaryClass = "hep-ph",
    doi = "10.1103/79cb-rssl",
    journal = "Phys. Rev. D",
    volume = "111",
    number = "12",
    pages = "123007",
    year = "2025"
}

@article{Costa:2025csj,
    author = "Costa, Francesco and Hoefken Zink, Jaime and Lucente, Michele and Pascoli, Silvia and Rosauro-Alcaraz, Salvador",
    title = "{Supercooled dark scalar phase transitions explanation of NANOGrav data}",
    eprint = "2501.15649",
    archivePrefix = "arXiv",
    primaryClass = "hep-ph",
    doi = "10.1016/j.physletb.2025.139634",
    journal = "Phys. Lett. B",
    volume = "868",
    pages = "139634",
    year = "2025"
}

@article{Jaeckel:2016jlh,
    author = "Jaeckel, Joerg and Khoze, Valentin V. and Spannowsky, Michael",
    title = "{Hearing the signal of dark sectors with gravitational wave detectors}",
    eprint = "1602.03901",
    archivePrefix = "arXiv",
    primaryClass = "hep-ph",
    reportNumber = "IPPP-16-12, DCPT-16-24",
    doi = "10.1103/PhysRevD.94.103519",
    journal = "Phys. Rev. D",
    volume = "94",
    number = "10",
    pages = "103519",
    year = "2016"
}

@article{vonHarling:2017yew,
    author = "von Harling, Benedict and Servant, Geraldine",
    title = "{QCD-induced Electroweak Phase Transition}",
    eprint = "1711.11554",
    archivePrefix = "arXiv",
    primaryClass = "hep-ph",
    reportNumber = "DESY-17-056",
    doi = "10.1007/JHEP01(2018)159",
    journal = "JHEP",
    volume = "01",
    pages = "159",
    year = "2018"
}

@article{Hambye:2018qjv,
    author = "Hambye, Thomas and Strumia, Alessandro and Teresi, Daniele",
    title = "{Super-cool Dark Matter}",
    eprint = "1805.01473",
    archivePrefix = "arXiv",
    primaryClass = "hep-ph",
    reportNumber = "ULB-TH/18-06, CERN-TH-2018-110, IFUP-TH/2018, ULB-TH-18-06",
    doi = "10.1007/JHEP08(2018)188",
    journal = "JHEP",
    volume = "08",
    pages = "188",
    year = "2018"
}

@article{Baldes:2018emh,
    author = "Baldes, Iason and Garcia-Cely, Camilo",
    title = "{Strong gravitational radiation from a simple dark matter model}",
    eprint = "1809.01198",
    archivePrefix = "arXiv",
    primaryClass = "hep-ph",
    reportNumber = "DESY 18-155, DESY-18-155",
    doi = "10.1007/JHEP05(2019)190",
    journal = "JHEP",
    volume = "05",
    pages = "190",
    year = "2019"
}

@article{Prokopec:2018tnq,
    author = "Prokopec, Tomislav and Rezacek, Jonas and {\'S}wie{\.z}ewska, Bogumi{\l}a",
    title = "{Gravitational waves from conformal symmetry breaking}",
    eprint = "1809.11129",
    archivePrefix = "arXiv",
    primaryClass = "hep-ph",
    doi = "10.1088/1475-7516/2019/02/009",
    journal = "JCAP",
    volume = "02",
    pages = "009",
    year = "2019"
}

@article{Kierkla:2023von,
    author = "Kierkla, Maciej and Swiezewska, Bogumila and Tenkanen, Tuomas V. I. and van de Vis, Jorinde",
    title = "{Gravitational waves from supercooled phase transitions: dimensional transmutation meets dimensional reduction}",
    eprint = "2312.12413",
    archivePrefix = "arXiv",
    primaryClass = "hep-ph",
    doi = "10.1007/JHEP02(2024)234",
    journal = "JHEP",
    volume = "02",
    pages = "234",
    year = "2024"
}

@article{Marzola:2017jzl,
    author = "Marzola, Luca and Racioppi, Antonio and Vaskonen, Ville",
    title = "{Phase transition and gravitational wave phenomenology of scalar conformal extensions of the Standard Model}",
    eprint = "1704.01034",
    archivePrefix = "arXiv",
    primaryClass = "hep-ph",
    doi = "10.1140/epjc/s10052-017-4996-1",
    journal = "Eur. Phys. J. C",
    volume = "77",
    number = "7",
    pages = "484",
    year = "2017"
}

@article{Mohamadnejad:2019vzg,
    author = "Mohamadnejad, Ahmad",
    title = "{Gravitational waves from scale-invariant vector dark matter model: Probing below the neutrino-floor}",
    eprint = "1907.08899",
    archivePrefix = "arXiv",
    primaryClass = "hep-ph",
    doi = "10.1140/epjc/s10052-020-7756-6",
    journal = "Eur. Phys. J. C",
    volume = "80",
    number = "3",
    pages = "197",
    year = "2020"
}

@article{Lewicki:2021xku,
    author = "Lewicki, Marek and Pujol{\`a}s, Oriol and Vaskonen, Ville",
    title = "{Escape from supercooling with or without bubbles: gravitational wave signatures}",
    eprint = "2106.09706",
    archivePrefix = "arXiv",
    primaryClass = "astro-ph.CO",
    doi = "10.1140/epjc/s10052-021-09669-6",
    journal = "Eur. Phys. J. C",
    volume = "81",
    number = "9",
    pages = "857",
    year = "2021"
}

@article{Schmitt:2024pby,
    author = "Schmitt, Daniel and Sagunski, Laura",
    title = "{QCD-sourced tachyonic phase transition in a supercooled Universe}",
    eprint = "2409.05851",
    archivePrefix = "arXiv",
    primaryClass = "hep-ph",
    doi = "10.1088/1475-7516/2025/02/075",
    journal = "JCAP",
    volume = "02",
    pages = "075",
    year = "2025"
}

@article{Datta:2022tab,
    author = "Datta, Satyabrata and Samanta, Rome",
    title = "{Gravitational waves-tomography of Low-Scale-Leptogenesis}",
    eprint = "2208.09949",
    archivePrefix = "arXiv",
    primaryClass = "hep-ph",
    doi = "10.1007/JHEP11(2022)159",
    journal = "JHEP",
    volume = "11",
    pages = "159",
    year = "2022"
}

@article{Datta:2023vbs,
    author = "Datta, Satyabrata and Samanta, Rome",
    title = "{Fingerprints of GeV scale right-handed neutrinos on inflationary gravitational waves and PTA data}",
    eprint = "2307.00646",
    archivePrefix = "arXiv",
    primaryClass = "hep-ph",
    doi = "10.1103/PhysRevD.108.L091706",
    journal = "Phys. Rev. D",
    volume = "108",
    number = "9",
    pages = "L091706",
    year = "2023"
}

@article{Giese:2020rtr,
    author = "Giese, Felix and Konstandin, Thomas and van de Vis, Jorinde",
    title = "{Model-independent energy budget of cosmological first-order phase transitions{\textemdash}A sound argument to go beyond the bag model}",
    eprint = "2004.06995",
    archivePrefix = "arXiv",
    primaryClass = "astro-ph.CO",
    reportNumber = "DESY-20-064",
    doi = "10.1088/1475-7516/2020/07/057",
    journal = "JCAP",
    volume = "07",
    number = "07",
    pages = "057",
    year = "2020"
}

@article{Ekstedt:2023sqc,
    author = "Ekstedt, Andreas and Gould, Oliver and Hirvonen, Joonas",
    title = "{BubbleDet: a Python package to compute functional determinants for bubble nucleation}",
    eprint = "2308.15652",
    archivePrefix = "arXiv",
    primaryClass = "hep-ph",
    doi = "10.1007/JHEP12(2023)056",
    journal = "JHEP",
    volume = "12",
    pages = "056",
    year = "2023"
}

@article{Hirvonen:2021zej,
    author = {Hirvonen, Joonas and L{\"o}fgren, Johan and Ramsey-Musolf, Michael J. and Schicho, Philipp and Tenkanen, Tuomas V. I.},
    title = "{Computing the gauge-invariant bubble nucleation rate in finite temperature effective field theory}",
    eprint = "2112.08912",
    archivePrefix = "arXiv",
    primaryClass = "hep-ph",
    reportNumber = "ACFI-T21-16, HIP-2021-45/TH, NORDITA 2021-111",
    doi = "10.1007/JHEP07(2022)135",
    journal = "JHEP",
    volume = "07",
    pages = "135",
    year = "2022"
}

@article{Lofgren:2021ogg,
    author = {L{\"o}fgren, Johan and Ramsey-Musolf, Michael J. and Schicho, Philipp and Tenkanen, Tuomas V. I.},
    title = "{Nucleation at Finite Temperature: A Gauge-Invariant Perturbative Framework}",
    eprint = "2112.05472",
    archivePrefix = "arXiv",
    primaryClass = "hep-ph",
    reportNumber = "ACFI-T21-15, HIP-2021-44/TH, NORDITA 2021-110",
    doi = "10.1103/PhysRevLett.130.251801",
    journal = "Phys. Rev. Lett.",
    volume = "130",
    number = "25",
    pages = "251801",
    year = "2023"
}

@article{Wainwright:2011qy,
    author = "Wainwright, Carroll and Profumo, Stefano and Ramsey-Musolf, Michael J.",
    title = "{Gravity Waves from a Cosmological Phase Transition: Gauge Artifacts and Daisy Resummations}",
    eprint = "1104.5487",
    archivePrefix = "arXiv",
    primaryClass = "hep-ph",
    reportNumber = "NPAC-11-04",
    doi = "10.1103/PhysRevD.84.023521",
    journal = "Phys. Rev. D",
    volume = "84",
    pages = "023521",
    year = "2011"
}

@article{Patel:2011th,
    author = "Patel, Hiren H. and Ramsey-Musolf, Michael J.",
    title = "{Baryon Washout, Electroweak Phase Transition, and Perturbation Theory}",
    eprint = "1101.4665",
    archivePrefix = "arXiv",
    primaryClass = "hep-ph",
    doi = "10.1007/JHEP07(2011)029",
    journal = "JHEP",
    volume = "07",
    pages = "029",
    year = "2011"
}

@article{Farakos:1994kx,
    author = "Farakos, K. and Kajantie, K. and Rummukainen, K. and Shaposhnikov, Mikhail E.",
    title = "{3-D physics and the electroweak phase transition: Perturbation theory}",
    eprint = "hep-ph/9404201",
    archivePrefix = "arXiv",
    reportNumber = "CERN-TH-6973-94, IUHET-273",
    doi = "10.1016/0550-3213(94)90173-2",
    journal = "Nucl. Phys. B",
    volume = "425",
    pages = "67--109",
    year = "1994"
}

@article{Braaten:1995cm,
    author = "Braaten, Eric and Nieto, Agustin",
    title = "{Effective field theory approach to high temperature thermodynamics}",
    eprint = "hep-ph/9501375",
    archivePrefix = "arXiv",
    reportNumber = "NUHEP-TH-95-2",
    doi = "10.1103/PhysRevD.51.6990",
    journal = "Phys. Rev. D",
    volume = "51",
    pages = "6990--7006",
    year = "1995"
}

@article{Kajantie:1995dw,
    author = "Kajantie, K. and Laine, M. and Rummukainen, K. and Shaposhnikov, Mikhail E.",
    title = "{Generic rules for high temperature dimensional reduction and their application to the standard model}",
    eprint = "hep-ph/9508379",
    archivePrefix = "arXiv",
    reportNumber = "CERN-TH-95-226, HU-TFT-95-50, IUHET-312",
    doi = "10.1016/0550-3213(95)00549-8",
    journal = "Nucl. Phys. B",
    volume = "458",
    pages = "90--136",
    year = "1996"
}

@article{Ekstedt:2022bff,
    author = "Ekstedt, Andreas and Schicho, Philipp and Tenkanen, Tuomas V. I.",
    title = "{DRalgo: A package for effective field theory approach for thermal phase transitions}",
    eprint = "2205.08815",
    archivePrefix = "arXiv",
    primaryClass = "hep-ph",
    reportNumber = "HIP-2022-11/TH, NORDITA 2022-030",
    doi = "10.1016/j.cpc.2023.108725",
    journal = "Comput. Phys. Commun.",
    volume = "288",
    pages = "108725",
    year = "2023"
}

@article{Dine:1992vs,
    author = "Dine, Michael and Leigh, Robert G. and Huet, Patrick and Linde, Andrei D. and Linde, Dmitri A.",
    title = "{Comments on the electroweak phase transition}",
    eprint = "hep-ph/9203201",
    archivePrefix = "arXiv",
    reportNumber = "SLAC-PUB-5740, SCIPP-92-06, SU-ITP-92-6",
    doi = "10.1016/0370-2693(92)90026-Z",
    journal = "Phys. Lett. B",
    volume = "283",
    pages = "319--325",
    year = "1992"
}

@article{Boyd:1993tz,
    author = "Boyd, C. Glenn and Brahm, David E. and Hsu, Stephen D. H.",
    title = "{Resummation methods at finite temperature: The Tadpole way}",
    eprint = "hep-ph/9304254",
    archivePrefix = "arXiv",
    reportNumber = "CALT-68-1858, HUTP-93-A011, EFI-93-22",
    doi = "10.1103/PhysRevD.48.4963",
    journal = "Phys. Rev. D",
    volume = "48",
    pages = "4963--4973",
    year = "1993"
}

@article{Curtin:2016urg,
    author = "Curtin, David and Meade, Patrick and Ramani, Harikrishnan",
    title = "{Thermal Resummation and Phase Transitions}",
    eprint = "1612.00466",
    archivePrefix = "arXiv",
    primaryClass = "hep-ph",
    reportNumber = "YITP-2016-48",
    doi = "10.1140/epjc/s10052-018-6268-0",
    journal = "Eur. Phys. J. C",
    volume = "78",
    number = "9",
    pages = "787",
    year = "2018"
}

@article{Curtin:2022ovx,
    author = "Curtin, David and Roy, Jyotirmoy and White, Graham",
    title = "{Gravitational waves and tadpole resummation: Efficient and easy convergence of finite temperature QFT}",
    eprint = "2211.08218",
    archivePrefix = "arXiv",
    primaryClass = "hep-ph",
    doi = "10.1103/PhysRevD.109.116001",
    journal = "Phys. Rev. D",
    volume = "109",
    number = "11",
    pages = "116001",
    year = "2024"
}

@article{Jackiw:1974cv,
    author = "Jackiw, R.",
    title = "{Functional evaluation of the effective potential}",
    doi = "10.1103/PhysRevD.9.1686",
    journal = "Phys. Rev. D",
    volume = "9",
    pages = "1686",
    year = "1974"
}

@article{Bodeker:2017cim,
    author = "Bodeker, Dietrich and Moore, Guy D.",
    title = "{Electroweak Bubble Wall Speed Limit}",
    eprint = "1703.08215",
    archivePrefix = "arXiv",
    primaryClass = "hep-ph",
    doi = "10.1088/1475-7516/2017/05/025",
    journal = "JCAP",
    volume = "05",
    pages = "025",
    year = "2017"
}

@article{Hoche:2020ysm,
    author = {H{\"o}che, Stefan and Kozaczuk, Jonathan and Long, Andrew J. and Turner, Jessica and Wang, Yikun},
    title = "{Towards an all-orders calculation of the electroweak bubble wall velocity}",
    eprint = "2007.10343",
    archivePrefix = "arXiv",
    primaryClass = "hep-ph",
    reportNumber = "FERMILAB-PUB-20-274-T",
    doi = "10.1088/1475-7516/2021/03/009",
    journal = "JCAP",
    volume = "03",
    pages = "009",
    year = "2021"
}

@article{Megevand:2016lpr,
    author = "Megevand, Ariel and Ramirez, Santiago",
    title = "{Bubble nucleation and growth in very strong cosmological phase transitions}",
    eprint = "1611.05853",
    archivePrefix = "arXiv",
    primaryClass = "astro-ph.CO",
    doi = "10.1016/j.nuclphysb.2017.03.009",
    journal = "Nucl. Phys. B",
    volume = "919",
    pages = "74--109",
    year = "2017"
}

@article{Cai:2017tmh,
    author = "Cai, Rong-Gen and Sasaki, Misao and Wang, Shao-Jiang",
    title = "{The gravitational waves from the first-order phase transition with a dimension-six operator}",
    eprint = "1707.03001",
    archivePrefix = "arXiv",
    primaryClass = "astro-ph.CO",
    reportNumber = "YITP-17-67",
    doi = "10.1088/1475-7516/2017/08/004",
    journal = "JCAP",
    volume = "08",
    pages = "004",
    year = "2017"
}

@misc{Liu:2025xvm,
    author = "Liu, Wei and Wu, Yongcheng",
    title = "{Testing Leptogenesis from Observable Gravitational Waves}",
    eprint = "2504.07819",
    archivePrefix = "arXiv",
    primaryClass = "hep-ph",
    reportNumber = "CPTNP-2025-001",
    month = "4",
    year = "2025"
}

@article{doi:10.1063/1.1338506,
	author = {Lorenz, Christian D. and Ziff, Robert M.},
	title = {{Precise determination of the critical percolation threshold for the three-dimensional “Swiss cheese” model using a growth algorithm}},
	journal = {The Journal of Chemical Physics},
	volume = {114},
	number = {8},
	pages = {3659-3661},
	year = {2001},
	doi = {10.1063/1.1338506},
	URL = {https://doi.org/10.1063/1.1338506},
	eprint = {https://doi.org/10.1063/1.1338506}
}

@article{LIN2018299,
	title = {Continuum percolation of porous media via random packing of overlapping cube-like particles},
	journal = {Theoretical and Applied Mechanics Letters},
	volume = {8},
	number = {5},
	pages = {299-303},
	year = {2018},
	issn = {2095-0349},
	doi = {https://doi.org/10.1016/j.taml.2018.05.007},
	url = {https://www.sciencedirect.com/science/article/pii/S209503491830196X},
	author = {Lin, Jianjun and Chen, Huisu}
}

@article{LI2020112815,
	title = {Numerical study for the percolation threshold and transport properties of porous composites comprising non-centrosymmetrical superovoidal pores},
	journal = {Computer Methods in Applied Mechanics and Engineering},
	volume = {361},
	pages = {112815},
	year = {2020},
	issn = {0045-7825},
	doi = {https://doi.org/10.1016/j.cma.2019.112815},
	url = {https://www.sciencedirect.com/science/article/pii/S0045782519307078},
	author = {Li, Mingqi and Chen, Huisu and Lin, Jianjun}
}

@phdthesis{2974136,
    author = "Mojahed, Martin Aria",
    title = "{Phenomenology of leptogenesis: From cosmological phase transitions to new mechanisms and experimental probes}",
    doi = "10.25358/openscience-13340",
    school = "Johannes Gutenberg University Mainz"
}

@misc{Ghosh:2025non,
    author = "Ghosh, Dilip Kumar and Ghoshal, Anish and Mukherjee, Koustav and Narendra, Nimmala and Okada, Nobuchika",
    title = "{Impact of Non-Thermal Leptogenesis with Early Matter Domination on Gravitational Waves from First-order Phase Transition}",
    eprint = "2508.02619",
    archivePrefix = "arXiv",
    primaryClass = "hep-ph",
    month = "8",
    year = "2025"
}
\end{document}